\documentclass[aps,preprint,amsfonts,amssymb,amsmath,eqsecnum,%
tightenlines,nofootinbib,showpacs,showkeys]{revtex4}

\newcommand{\cN}{\mathcal{N}}
\newcommand{\cP}{\mathcal{P}}
\newcommand{\cT}{\mathcal{T}}
\newcommand{\cV}{\mathcal{V}}
\newcommand{\bH}{\boldsymbol{H}}
\newcommand{\bQ}{\boldsymbol{Q}}
\newcommand{\fF}{\mathfrak{F}}
\newcommand{\bbN}{\mathbb{N}}

\newcommand{\rmd}{\mathrm{d}}
\newcommand{\del}{\partial}
\newcommand{\sV}{\mathsf{V}}

\begin{document}



%
%

\title{Parasupersymmetry and $\cN$-fold Supersymmetry in Quantum
 Many-Body Systems II. Third Order}
\author{Toshiaki Tanaka}
\email{ttanaka@mail.ncku.edu.tw}
\affiliation{Department of Physics, National Cheng-Kung University,\\
 Tainan 701, Taiwan, R.O.C.\\
 National Center for Theoretical Sciences, Taiwan, R.O.C.}


\begin{abstract}

Based on the general formalism of parafermionic algebra and
parasupersymmetry proposed previously by us, we explicitly
construct third-order parafermionic algebra and multiplication
law, and then realize third-order parasupersymmetric quantum
systems. We find some novel features in the third-order, namely,
the emergence of a fermionic degree of freedom and of a generalized
parastatistics. We show that for one-body cases the generalized
Rubakov--Spiridonov model can be constructed also in our framework
and find that it admits a generalized $3$-fold superalgebra. We also
find that a three-body system can have third-order parasupersymmetry
where three independent supersymmetries are folded. In both cases, we
also investigate the new concept of quasi-parasupersymmetry introduced
by us and find that those of order $(3,3)$ are indeed realized under
less restrictive conditions than (ordinary) parasupersymmetric cases.

\end{abstract}






\maketitle

\section{Introduction}
\label{sec:intro}

In our previous article \cite{Ta07a}, we proposed the general
formalism of parafermionic algebra and parasupersymmetric quantum
systems (for an extensive bibliography, see the references cited
therein) without recourse to any specific matrix
representation and any kind of deformed oscillator algebra so that
we can investigate and discuss general aspects of them. Within
the formalism, we showed generically that every parasupersymmetric
system of order $p$ consists of $\cN$-fold supersymmetric pairs of
component Hamiltonians with $\cN\leq q$ and thus they have isospectral
property and weak quasi-solvability. The implication of the result
in view of higher-dimensional quantum theories is that
parasupersymmetric quantum field theory (cf. Refs.~\cite{FV91,BD93b})
as well as weak supersymmetric one \cite{Sm03}, if exists as a
consistent theory, would satisfy some kinds of perturbative
non-renormalization theorems, as in the case of ordinary supersymmetric
quantum field theory, since quasi-solvability is a one-dimensional
analog and, in a sense, a generalization of them~\cite{AST01b}.
To illustrate how the formalism works, in the article \cite{Ta07a}
we constructed explicitly second-order parafermionic algebra based
on solely the postulates of the formalism and some second-order
parasupersymmetric quantum systems.

In this article, we proceed with the study of parasupersymmetry based
on the previous formalism in Ref.~\cite{Ta07a} and focus on, in
particular, third-order case. First of all, we would like to check
whether the postulates proposed in Ref.~\cite{Ta07a} work consistently
not only in the second-order case but also in higher-order cases.
Second, we would like to observe what kind of novel features can arise
when the parafermionic order increases. In particular, the new concept
of quasi-parasupersymmetry did not produce any new results in the
second-order case, and thus it is interesting to see the situation
in higher-order cases. Third, the direct construction of parafermionic
algebra based on the postulates must become more cumbersome as the
parafermionic order increases. Hence, we would like to acquire more
general features in lower-order cases so that we would be able to
establish an inductive construction of parafermionic algebra and
multiplication law for arbitrary order. In fact, it would be almost
impossible to guess the form of them only by the knowledge of the
second-order case.

Motivated by the above purposes, in this article we first construct
third-order parafermionic algebra and multiplication law by following
a similar way of the derivation of the second-order ones in
Ref.~\cite{Ta07a}. We confirm that the postulates of the formalism
work consistently in the third-order case too, and find some novel
features in the third-order parafermionic algebra.
We then construct some quantum systems with third-order
parasupersymmetry and quasi-parasupersymmetry. For one-body
systems, we find that the generalized Rubakov--Spiridonov model
\cite{RS88,Kh92,Kh93} of order $3$ is realized also in our framework
and further show that it admits a generalized $3$-fold superalgebra.
Furthermore, we find that a three-body system also admits third-order
parasupersymmetry where three independent supersymmetries are folded.
In both cases, we also investigate quasi-parasupersymmetry and find
that order $(3,2)$ cases are identical with (ordinary) parasupersymmetric
ones while order $(3,3)$ cases are indeed less restrictve than them due
to the lack of a constraint on one of the component Hamiltonians.

We organize the article as follows. In the next section, we review
the general formalism of parafermionic algebra and
(quasi-)parasupersymmetry in quantum mechanical systems developed
in Ref.~\cite{Ta07a}. In Section~\ref{sec:palg3}, we construct
third-order parafermionic algebra solely based on the postulates in
the formalism and derive the multiplication law. There some novel
features of the third-order case are pointed out. In
Section~\ref{sec:3psqs}, using the parafermionic algebra and
multiplication law of order $3$, we investigate and construct
two different parasupersymmetric quantum systems of order $3$.
The one consists of one-body Hamiltonians and is identical to the model
in Refs.~\cite{Kh92,Kh93}. We also show that this system admits
a generalized $3$-fold superalgebra in an analogous way
the second-order Rubakov--Spiridonov model admits a generalized
$2$-fold superalgebra shown in Ref.~\cite{Ta07a}.
The other third-order model consists of three-body Hamiltonians and
has three independent supersymmetric structures. Furthermore, we
investigate quasi-parasupersymmetry in each of the model. Finally,
we summarize and discuss the results in Section~\ref{sec:discus}.

\section{Review of (Quasi-)Parasupersymmetry}
\label{sec:rev}

First of all, let us define parafermionic algebra of order $p(\in\bbN)$.
It is an associative algebra composed of the identity operator $I$
and two parafermionic operators $\psi^{-}$ and $\psi^{+}$ of order
$p$ which satisfy the nilpotency:
\begin{align}
(\psi^{-})^{p}\neq 0,\quad (\psi^{+})^{p}\neq 0,\qquad
 (\psi^{-})^{p+1}=(\psi^{+})^{p+1}=0.
\label{eq:nilpo}
\end{align}
Hence, we immediately have $2p+1$ non-zero elements
$\{I,\psi^{-},\dots,(\psi^{-})^{p},\psi^{+},\dots,(\psi^{+})^{p}\}$.
We call them the \emph{fundamental} elements of parafermionic
algebra of order $p$. Parafermionic algebra is characterized by
anti-commutation relation $\{A,B\}=AB+BA$ and commutation relation
$[A,B]=AB-BA$ among the fundamental elements. As a postulate we
impose the following relation for arbitrary order $p$: 
\begin{align}
\{\psi^{-},\psi^{+}\}+\{(\psi^{-})^{2},(\psi^{+})^{2}\}+\dots
 +\{(\psi^{-})^{p},(\psi^{+})^{p}\}=pI.
\label{eq:postu}
\end{align}
We shall next define parafermionic Fock spaces $\sV_{p}$ of order $p$
on which the parafermionic operators act. The latter space is $(p+1)$
dimensional and its $p+1$ bases $|k\rangle$ ($k=0,\dots,p$)
are defined by
\begin{align}
\psi^{-}|0\rangle=0,\quad |k\rangle=(\psi^{+})^{k}|0\rangle,
 \quad\psi^{-}|k\rangle=|k-1\rangle\quad (k=1,\dots,p).
\label{eq:defpfs}
\end{align}
That is, $\psi^{-}$ and $\psi^{+}$ act as annihilation and
creation operators of parafermions, respectively.
The state $|0\rangle$ is called the parafermionic \emph{vacuum}.
The subspace spanned by each state $|k\rangle$ ($k=0,\dots,p$) is
called the $k$-parafermionic subspace and is denoted by
$\sV_{p}^{(k)}$. The adjoint vector $\langle k|$ of each $|k
\rangle$ is introduced as a linear operator which maps every
vector in $\sV_{p}$ into a complex number as follows:
\begin{align}
\langle k|l\rangle=\langle 0|(\psi^{-})^{k}|l\rangle,\quad
\langle 0|l\rangle=\langle 0|(\psi^{+})^{l}|0\rangle
 =\delta_{0,l}\quad (k,l=0,\dots,p).
\label{eq:defadv}
\end{align}
By the definitions (\ref{eq:defpfs}) and (\ref{eq:defadv}),
we immediately have a bi-orthogonal relation:
\begin{align}
\langle k|l\rangle=\delta_{k,l}\quad (k,l=0,\dots,p).
\end{align}
We can now define a set of projection operators
$\Pi_{k}:\sV_{p} \to\sV_{p}^{(k)}$ ($k=0,\dots,p$) which satisfy
\begin{align}
\Pi_{k}|l\rangle=\delta_{k,l}|k\rangle,\qquad\Pi_{k}\Pi_{l}=
 \delta_{k,l}\Pi_{k},\qquad\sum_{k=0}^{p}\Pi_{k}=I.
\label{eq:defpo}
\end{align}
From the definitions (\ref{eq:defpfs}) and (\ref{eq:defpo}), we obtain
\begin{align}
\Pi_{k+1}\psi^{+}=\psi^{+}\Pi_{k},\qquad
 \psi^{-}\Pi_{k+1}=\Pi_{k}\psi^{-},
\label{eq:Pipsi}
\end{align}
where and hereafter we put $\Pi_{k}\equiv 0$ for all $k<0$ and $k>p$.

We now come back to the parafermionic algebra. Apparently, the
relations (\ref{eq:nilpo}) and (\ref{eq:postu}) are not sufficient
for the determination of the full algebra. To determine other
multiplication relations we impose the following postulates:
\begin{enumerate}
\item First, the algebra must be consistent with
 Eq.~(\ref{eq:defpfs}).
 This requirement is indispensable for defining consistently
 the parafermionic Fock space $\sV_{p}$.
\item Every projection operator $\Pi_{k}$ ($k=0,\dots,p$) can be
 expressed as a polynomial of the fundamental elements of the
 corresponding order $p$ so that the algebra is consistent with
 the definition (\ref{eq:defpo}).
\item Every product of three fundamental elements can be expressed
 as a polynomial of at most second-degree in the fundamental elements.
 These formulas are called the \emph{multiplication law}. In particular,
 we assume that the following relations
 \begin{align}
 \psi^{-}\psi^{+}\psi^{-}=\psi^{-},\qquad
  \psi^{+}\psi^{-}\psi^{+}=\psi^{+}.
 \label{eq:mult1}
 \end{align}
 hold for parafermionic operators of \emph{any} order $p$.
 As a consequence of this assumption, we immediately obtain for all
 $m,n\in\bbN$
 \begin{align}
 (\psi^{-})^{m}\psi^{+}(\psi^{-})^{n}=(\psi^{-})^{m+n-1},\quad
 (\psi^{+})^{m}\psi^{-}(\psi^{+})^{n}=(\psi^{+})^{m+n-1},
 \label{eq:mult2}
 \end{align}
 which also hold for arbitrary order.
\item We also assume that the following relations hold for arbitrary
 order:
 \begin{align}
 (\psi^{-})^{p}\psi^{+}\Pi_{p-1}=(\psi^{-})^{p-1}\Pi_{p-1},\qquad
 \psi^{+}(\psi^{-})^{p}\Pi_{p}=(\psi^{-})^{p-1}\Pi_{p}.
 \label{eq:mult3}
 \end{align}
\end{enumerate}
We note that every polynomial composed of the fundamental elements
can be reduced to a polynomial of at most second-degree in them
as a consequence of the third postulate and the associativity.
Hence, together with the second postulate it means in particular
every projection operators must be expressed as a polynomial of
second-degree in the fundamental elements.

Finally, we introduce the quantity of parafermionic \emph{degree}
of operators as follows:
\begin{align}
\deg I=0,\qquad\deg\psi^{+}=1,\qquad\deg\psi^{-}=p,\\
\deg AB\equiv\deg A+\deg B\pmod{p+1}.
\end{align}
For example, $\deg(\psi^{+})^{k}=k$ and $\deg(\psi^{-})^{k}=p+1-k$
($k=1,\dots,p$).\\

Parasupersymmetry of order 2 in quantum mechanics was first introduced
by Rubakov and Spiridonov \cite{RS88} and was later generalized
to arbitrary order independently by Tomiya~\cite{To92} and by
Khare~\cite{Kh92}. A different formulation for order 2 was proposed
by Beckers and Debergh \cite{BD90a} and a generalization of the latter
to arbitrary order was attempted by Chenaghlou and Fakhri \cite{CF03}.
Thus, we call them RSTK and BDCF formalism, respectively.
To define a $p$th-order parasupersymmetric system, we first introduce
a pair of parasupercharges $\bQ^{\pm}$ of order $p$ which satisfy
\begin{align}
(\bQ^{-})^{p}\neq 0,\quad(\bQ^{+})^{p}\neq 0,\quad
 (\bQ^{-})^{p+1}=(\bQ^{+})^{p+1}=0.
\label{eq:pfsc1}
\end{align}
A system $\bH$ is said to have \emph{parasupersymmetry of order $p$} if
it commutes with the parasupercharges of order $p$
\begin{align}
[\bQ^{-},\bH]=[\bQ^{+},\bH]=0,
\label{eq:pfsc2}
\end{align}
and satisfies the non-linear relations in the RSTK formalism
\begin{align}
\sum_{k=0}^{p}(\bQ^{-})^{p-k}\bQ^{+}(\bQ^{-})^{k}
 =C_{p}(\bQ^{-})^{p-1}\bH,\quad
\sum_{k=0}^{p}(\bQ^{+})^{p-k}\bQ^{-}(\bQ^{+})^{k}
 =C_{p}\bH(\bQ^{+})^{p-1},
\label{eq:pfsc3}
\end{align}
or in the BDCF formalism
\begin{subequations}
\label{eq:pfsc4}
\begin{align}
\underbrace{[\bQ^{-},\cdots,[\bQ^{-}}_{(p-1)\text{ times}},
 [\bQ^{+},\bQ^{-}]]\cdots]&=(-1)^{p}C_{p}(\bQ^{-})^{p-1}\bH,\\
\underbrace{[\bQ^{-},\cdots,[\bQ^{+}}_{(p-1)\text{ times}},
 [\bQ^{-},\bQ^{+}]]\cdots]&=C_{p}\bH(\bQ^{+})^{p-1},
\end{align}
\end{subequations}
where $C_{p}$ is a constant.
An apparent drawback of the BDCF formalism is that the relations
(\ref{eq:pfsc4}) do not reduce to the ordinary supersymmetric
anti-commutation relation $\{\bQ^{-},\bQ^{+}\}=C_{1}\bH$ when $p=1$,
in contrast to the RSTK relation (\ref{eq:pfsc3}). For this reason, we
discard the BDCF formalism in this article though its defect may be
amended by, e.g., replacing all the commutators in (\ref{eq:pfsc4})
by anti-commutators, graded commutators $[A,B\}=AB-
(-1)^{\deg A\cdot\deg B}BA$, and so on (see also the third paragraph
in Section~\ref{sec:discus}).

An immediate consequence of the commutativity (\ref{eq:pfsc2}) is that
each $n$th-power of the parasupercharges ($2\leq n\leq p$) also commutes
with the system $\bH$
\begin{align}
[(\bQ^{-})^{n},\bH]=[(\bQ^{+})^{n},\bH]=0\quad(2\leq n\leq p).
\label{eq:pfsc5}
\end{align}
Hence, every parasupersymmetric system $\bH$ satisfying (\ref{eq:pfsc2})
always has $2p$ conserved charges.

To realize parasupersymmetry in quantum mechanical systems, we usually
consider a vector space $\fF\times\sV_{p}$ where $\fF$ is a linear space
of complex functions such as the Hilbert space $L^{2}$ in Hermitian
quantum theory and the Krein space $L_{\cP}^{2}$ in
$\cP\cT$-symmetric quantum theory \cite{Ta06b,Ta06d}. A parafermionic
quantum system $\bH$ is introduced by
\begin{align}
\bH=\sum_{k=0}^{p}H_{k}\Pi_{k},
\label{eq:pfqs}
\end{align}
where $H_{k}$ ($k=0,\dots,p$) are scalar Hamiltonians of $p$ variables
acting on $\fF$:
\begin{align}
H_{k}=-\frac{1}{2}\sum_{i=1}^{p}\frac{\partial_{i}^{2}}{\partial
 q_{i}^{2}}+V_{k}(q_{1},\cdots,q_{p})\quad (k=0,\dots,p).
\label{eq:Schro}
\end{align}
Two parasupercharges $\bQ^{\pm}$ are defined by
\begin{align}
\bQ^{-}=\sum_{k=0}^{p}Q_{k}^{-}\psi^{-}\Pi_{k},
 \qquad\bQ^{+}=\sum_{k=0}^{p}Q_{k}^{+}\Pi_{k}\psi^{+},
\label{eq:pfsc}
\end{align}
where $Q_{k}^{+}$ ($k=0,\dots,p$) are first-order linear operators
acting on $\fF$
\begin{align}
Q_{k}^{+}=\sum_{i=1}^{p}w_{k,i}(q_{1},\dots,q_{p})\frac{\partial}{
 \partial q_{i}}+W_{k}(q_{1},\dots,q_{p})\quad(k=0,\dots,p),
\label{eq:compQ}
\end{align}
and for each $k$ $Q_{k}^{-}$ is given by a certain `adjoint'
of $Q_{k}^{+}$, e.g., the (ordinary) adjoint $Q_{k}^{-}=
(Q_{k}^{+})^{\dagger}$ in the Hilbert space $L^{2}$,
the $\cP$-adjoint $Q_{k}^{-}=\cP (Q_{k}^{+})^{\dagger}\cP$ in
the Krein space $L_{\cP}^{2}$, and so on. For all $k\leq 0$ and
$k>p$ we put $Q_{k}^{\pm}\equiv 0$.
When $p=1$, the triple $(\bH,\bQ^{-},\bQ^{+})$ defined in
Eqs.~(\ref{eq:pfqs}) and (\ref{eq:pfsc}) becomes
\begin{align}
\bH=H_{0}\psi^{-}\psi^{+}+H_{1}\psi^{+}\psi^{-},\quad
\bQ^{-}=Q_{1}^{-}\psi^{-},\quad\bQ^{+}=Q_{1}^{+}\psi^{+},
\end{align}
and thus reduces to an ordinary supersymmetric quantum mechanical
system \cite{Wi81}. The non-linear relation (\ref{eq:pfsc3}) together
with the nilpotency (\ref{eq:pfsc1}) for $p=1$ are just the
anti-commutation relations between supercharges
\begin{align}
\{\bQ^{\pm},\bQ^{\pm}\}=0,\qquad\{\bQ^{-},\bQ^{+}\}=C_{1}\bH.
\end{align}
Hence, the parasupersymmetric quantum systems defined by
Eqs.~(\ref{eq:pfsc1})--(\ref{eq:compQ}) provide a natural
generalization of ordinary supersymmetric quantum mechanics.

It is easy to check that the parasupercharges $\bQ^{\pm}$ defined by
Eq.~(\ref{eq:pfsc}) already satisfy the nilpotency (\ref{eq:pfsc1}) and
that the commutativity (\ref{eq:pfsc2}) is satisfied if and only if
\begin{align}
H_{k-1}Q_{k}^{-}=Q_{k}^{-}H_{k},\quad 
 Q_{k}^{+}H_{k-1}=H_{k}Q_{k}^{+},\quad\forall k=1,\dots,p.
\label{eq:inter}
\end{align}
That is, each pair of $H_{k-1}$ and $H_{k}$ must satisfy the
intertwining relations with respect to $Q_{k}^{-}$ and $Q_{k}^{+}$.
Similarly, the commutativity (\ref{eq:pfsc5}) implies that any pair of
$H_{k-n}$ and $H_{k}$ ($1\leq n\leq k\leq p$) satisfies
\begin{subequations}
\label{eq:Nfold}
\begin{align}
H_{k-n}Q_{k-n+1}^{-}\cdots Q_{k-1}^{-}Q_{k}^{-}
 =Q_{k-n+1}^{-}\dots Q_{k-1}^{-}Q_{k}^{-}H_{k},\\
Q_{k}^{+}Q_{k-1}^{+}\dots Q_{k-n+1}^{+}H_{k-n}
 =H_{k}Q_{k}^{+}Q_{k-1}^{+}\cdots Q_{k-n+1}^{+},
\end{align}
\end{subequations}
which means that $H_{k-n}$ and $H_{k}$ constitute a pair of
$\cN$-fold supersymmetry with $\cN=n$. The relations (\ref{eq:Nfold})
can be also derived by repeated applications of Eq.~(\ref{eq:inter}).
Since $\cN$-fold supersymmetry is essentially equivalent to weak
quasi-solvability \cite{AST01b,Ta03a}, parasupersymmetric quantum
systems also possess weak quasi-solvability. To see the structure
of weak quasi-solvability in the parasupersymmetric system $\bH$
more precisely, let us first define
\begin{align}
\cV_{n,k}^{-}=\ker (Q_{k-n+1}^{-}\cdots Q_{k}^{-}),\quad
 \cV_{n,k}^{+}=\ker (Q_{k}^{+}\cdots Q_{k-n+1}^{+})\quad
 (1\leq n\leq k\leq p).
\label{eq:defVnk}
\end{align}
By the definition (\ref{eq:defVnk}), the vector spaces
$\cV_{n,k}^{\pm}$ for each fixed $k$ are related as
\begin{align}
\cV_{1,k}^{-}\subset\cV_{2,k}^{-}\subset\cdots\subset
 \cV_{k,k}^{-},\quad
\cV_{1,k}^{+}\subset\cV_{2,k}^{+}\subset\cdots\subset
 \cV_{k,k}^{+}.
\label{eq:flag}
\end{align}
On the other hand, it is evident from the intertwining relations
(\ref{eq:Nfold}) that each Hamiltonian $H_{k}$ ($0\leq k\leq p$)
preserves vector spaces as follows:
\begin{subequations}
\label{eq:invsp}
\begin{align}
H_{k}\cV_{n,k}^{-}\subset\cV_{n,k}^{-}\quad(1\leq n\leq k),\\
H_{k}\cV_{n,k+n}^{+}\subset\cV_{n,k+n}^{+}\quad(1\leq n\leq p-k).
\end{align}
\end{subequations}
From Eqs.~(\ref{eq:flag}) and (\ref{eq:invsp}), the largest
space preserved by each $H_{k}$ ($0\leq k\leq p$) is given by
\begin{align}
\cV_{k,k}^{-}+\cV_{p-k,p}^{+}\quad(0\leq k\leq p).
\label{eq:lstinv}
\end{align}
Needless to say, each Hamiltonian $H_{k}$ preserves the two spaces
in Eq.~(\ref{eq:lstinv}) separately.
The intertwining relations (\ref{eq:inter}) and (\ref{eq:Nfold})
ensure that all the component Hamiltonians $H_{k}$ ($k=0,\dots,p$) of
the system $\bH$ are isospectral outside the sectors $\cV_{n,k}^{\pm}$
($1\leq n\leq k\leq p$). The spectral degeneracy of $\bH$ in these
sectors depends on the form of each component of the parasupercharges,
$Q_{k}^{\pm}$ ($k=1,\dots,p$), and its structure can be very complicated
even in the case of second-order, see e.g. Refs.~\cite{Mo96a,Mo97}.

In addition to those `power-type' symmetries, every parasupersymmetric
quantum system $\bH$ defined in Eq.~(\ref{eq:pfqs}) can have
`discrete-type' ones. The conserved charges of this type are given by
\begin{align}
\bQ_{\{n\}}^{\pm}=[\{(\psi^{-})^{n},(\psi^{+})^{n}\},\bQ^{\pm}],
 \quad\bQ_{[n]}^{\pm}=[[(\psi^{-})^{n},(\psi^{+})^{n}],\bQ^{\pm}]
 \quad (n=1,\ldots,p).
\label{eq:discc}
\end{align}
It follows from Jacobi identity that they indeed commute with $\bH$:
\begin{align}
[\bQ_{\{n\}}^{\pm},\bH]=[\bQ_{[n]}^{\pm},\bH]=0\quad
 (n=1,\dots,p).
\end{align}
We note, however, that they are in general not linearly independent
and we cannot determine the number of linearly independent conserved
charges without the knowledge of parafermionic algebra of each order.

The non-linear relations (\ref{eq:pfsc3}) can be also calculated in
a similar way. The first non-linear relation in Eq.~(\ref{eq:pfsc3})
is satisfied if and only if the following two identities hold:
\begin{subequations}
\label{eq:parac}
\begin{align}
Q_{1}^{-}\cdots Q_{p}^{-}Q_{p}^{+}+\sum_{k=1}^{p-1}Q_{1}^{-}
 \cdots Q_{p-k}^{-}Q_{p-k}^{+}Q_{p-k}^{-}\cdots Q_{p-1}^{-}
 =C_{p}Q_{1}^{-}\cdots Q_{p-1}^{-}H_{p-1},
\label{eq:parac1}\\
\sum_{k=1}^{p-1}Q_{2}^{-}\cdots Q_{p-k+1}^{-}Q_{p-k+1}^{+}
 Q_{p-k+1}^{-}\cdots Q_{p}^{-}+Q_{1}^{+}Q_{1}^{-}\cdots Q_{p}^{-}
 =C_{p}Q_{2}^{-}\cdots Q_{p}^{-}H_{p}.
\end{align}
\end{subequations}
The conditions for the second non-linear relation in
Eq.~(\ref{eq:pfsc3}) are apparently given by the `adjoint' of
Eqs.~(\ref{eq:parac}).\\

With a given pair of parasupercharges $\bQ^{\pm}$ of order $p$
which satisfy the nilpotency (\ref{eq:pfsc1}), a system $\bH$
is said to have \emph{quasi-parasupersymmetry of order} $(p,q)$
if there exists a natural number $q$ ($1\leq q\leq p$) such
that $(\bQ^{\pm})^{q}$ commutes with $\bH$ and the non-linear
constraint (\ref{eq:pfsc3}) is satisfied. That is, it is
characterized by the following algebraic relations:
\begin{align}
(\bQ^{-})^{p}\neq 0,\quad (\bQ^{+})^{p}\neq 0,\quad
 (\bQ^{-})^{p+1}=(\bQ^{+})^{p+1}=0,\\[10pt]
[(\bQ^{-})^{q},\bH]=[(\bQ^{+})^{q},\bH]=0\quad (1\leq q\leq p),\\
\sum_{k=0}^{p}(\bQ^{-})^{p-k}\bQ^{+}(\bQ^{-})^{k}
 =C_{p}(\bQ^{-})^{p-1}\bH,\quad
\sum_{k=0}^{p}(\bQ^{+})^{p-k}\bQ^{-}(\bQ^{+})^{k}
 =C_{p}\bH(\bQ^{+})^{p-1}.
\label{eq:qpfsc3}
\end{align}
By definition, quasi-parasupersymmetry of order $(p,q)$ reduces
to (ordinary) parasupersymmetry when $q=1$. Thus, it can be
regarded as a generalization of parasupersymmetry. A key
ingredient of this new symmetry is that the commutativity
$[(\bQ^{\pm})^{n},\bH]=0$ for $n<q$ is not necessarily
fulfilled in contrast to (ordinary) parasupersymmetry.
As a consequence, only the less restrictive $q$th-order intertwining
relations (\ref{eq:Nfold}) with $n=q$ should be satisfied between
every pair of $H_{k-q}$ and $H_{k}$ in the case of
quasi-parasupersymmetry of order $(p,q)$. The `power-type' conserved
charges in this case are apparently given by
\begin{align}
[(\bQ^{-})^{qn},\bH]=[(\bQ^{+})^{qn},\bH]=0\quad (2\leq n\leq
 [{\textstyle\frac{p}{q}}]),
\end{align}
where $[x]$ is the maximum integer which does not exceed $x$,
and thus the number of conserved charges is reduced to
$2[\frac{p}{q}]$. It is evident that parasupersymmetry of order
$p$ always implies quasi-parasupersymmetry of order $(p,q)$ for
all $q=1,\dots,p$. The `discrete-type'
conserved charges (cf. Eq.~(\ref{eq:discc})) are similarly defined.

\section{Parafermionic Algebra of Order $3$}
\label{sec:palg3}

In this section, we shall construct parafermionic algebra of order 3
based on the postulates in Section~\ref{sec:rev}. The starting
point is the relations (\ref{eq:nilpo}) and (\ref{eq:postu}) for $p=3$:
\begin{align}
(\psi^{-})^{4}=(\psi^{+})^{4}=0,
\label{eq:3alg0}\\
\{\psi^{-},\psi^{+}\}+\{(\psi^{-})^{2},(\psi^{+})^{2}\}
 +\{(\psi^{-})^{3},(\psi^{+})^{3}\}=3I.
\label{eq:3alg1}
\end{align}
We note that any formula derived from Eqs.~(\ref{eq:3alg0}) and
(\ref{eq:3alg1}) also hold when all the indices of $+$ and $-$ in
the formula are simultaneously interchanged since the original algebra
(\ref{eq:3alg0}) and (\ref{eq:3alg1}) is invariant under
the interchange of $+$ and $-$. 
First, multiplying (\ref{eq:3alg1}) by three $\psi^{-}$s as
$(\psi^{-})^{3}\times$(\ref{eq:3alg1}), $(\psi^{-})^{2}\times$
(\ref{eq:3alg1})$\times\psi^{-}$, $\psi^{-}\times(\ref{eq:3alg1})\times
(\psi^{-})^{2}$, and (\ref{eq:3alg1})$\times(\psi^{-})^{3}$, and applying
the assumption (\ref{eq:mult2}) and the nilpotency (\ref{eq:3alg0}),
we immediately obtain
\begin{subequations}
\label{eq:3alg2}
\begin{align}
(\psi^{-})^{3}(\psi^{+})^{3}(\psi^{-})^{3}=(\psi^{-})^{3}(\psi^{+})^{2}
 (\psi^{-})^{2}=(\psi^{-})^{2}(\psi^{+})^{2}(\psi^{-})^{3}
 =(\psi^{-})^{3},\\
(\psi^{-})^{3}(\psi^{+})^{2}(\psi^{-})^{3}=0,
\end{align}
\end{subequations}
and thus their sign-interchanged relations
\begin{subequations}
\label{eq:3alg2'}
\begin{align}
(\psi^{+})^{3}(\psi^{-})^{3}(\psi^{+})^{3}=(\psi^{+})^{3}(\psi^{-})^{2}
 (\psi^{+})^{2}=(\psi^{+})^{2}(\psi^{-})^{2}(\psi^{+})^{3}
 =(\psi^{+})^{3},\\
(\psi^{+})^{3}(\psi^{-})^{2}(\psi^{+})^{3}=0.
\end{align}
\end{subequations}
Next, multiplying (\ref{eq:3alg1}) by $(\psi^{+})^{2}$ and
$(\psi^{-})^{2}$ as $(\psi^{+})^{2}(\psi^{-})^{2}\times$(\ref{eq:3alg1}),
$(\psi^{+})^{2}\times$(\ref{eq:3alg1})$\times(\psi^{-})^{2}$, and
(\ref{eq:3alg1})$\times(\psi^{+})^{2}(\psi^{-})^{2}$, and applying
(\ref{eq:mult2}), (\ref{eq:3alg0}), and the formulas
(\ref{eq:3alg2})--(\ref{eq:3alg2'}), we have
\begin{subequations}
\label{eq:3alg3}
\begin{align}
(\psi^{+})^{2}(\psi^{-})^{3}\psi^{+}+(\psi^{+})^{2}(\psi^{-})^{2}
 (\psi^{+})^{2}(\psi^{-})^{2}+(\psi^{+})^{3}(\psi^{-})^{3}
 &=2(\psi^{+})^{2}(\psi^{-})^{2},\\
(\psi^{+})^{3}(\psi^{-})^{3}+(\psi^{+})^{2}(\psi^{-})^{2}(\psi^{+})^{2}
 (\psi^{-})^{2}+(\psi^{+})^{2}(\psi^{-})^{3}(\psi^{+})^{3}(\psi^{-})^{2}
 &=2(\psi^{+})^{2}(\psi^{-})^{2},\\
\psi^{-}(\psi^{+})^{3}(\psi^{-})^{2}+(\psi^{+})^{2}(\psi^{-})^{2}
 (\psi^{+})^{2}(\psi^{-})^{2}+(\psi^{+})^{3}(\psi^{-})^{3}
 &=2(\psi^{+})^{2}(\psi^{-})^{2}.
\end{align}
\end{subequations}
From the above set of equations, we obtain
\begin{align}
(\psi^{+})^{2}(\psi^{-})^{3}\psi^{+}=(\psi^{+})^{2}(\psi^{-})^{3}
 (\psi^{+})^{3}(\psi^{-})^{2}=\psi^{-}(\psi^{+})^{3}(\psi^{-})^{2}.
\label{eq:3alg4}
\end{align}
Multiplying (\ref{eq:3alg4}) by $\psi^{+}$ from left or by $\psi^{-}$
from right, and applying the formulas (\ref{eq:3alg2})--(\ref{eq:3alg2'}),
we get
\begin{align}
(\psi^{+})^{3}(\psi^{-})^{3}\psi^{+}=(\psi^{+})^{3}(\psi^{-})^{2},\qquad
 \psi^{-}(\psi^{+})^{3}(\psi^{-})^{3}=(\psi^{+})^{2}(\psi^{-})^{3},
\label{eq:3alg5}
\end{align}
and thus their sign-interchanged ones
\begin{align}
(\psi^{-})^{3}(\psi^{+})^{3}\psi^{-}=(\psi^{-})^{3}(\psi^{+})^{2},\qquad
 \psi^{+}(\psi^{-})^{3}(\psi^{+})^{3}=(\psi^{-})^{2}(\psi^{+})^{3}.
\label{eq:3alg5'}
\end{align}
As a consequence of the formulas (\ref{eq:3alg5})--(\ref{eq:3alg5'}),
we immediately have
\begin{align}
(\psi^{+})^{3}(\psi^{-})^{3}(\psi^{+})^{2}=(\psi^{+})^{3}(\psi^{-})^{2}
 \psi^{+},\qquad (\psi^{-})^{2}(\psi^{+})^{3}(\psi^{-})^{3}
 =\psi^{-}(\psi^{+})^{2}(\psi^{-})^{3},
\label{eq:3alg6}\\
(\psi^{-})^{3}(\psi^{+})^{3}(\psi^{-})^{2}=(\psi^{-})^{3}(\psi^{+})^{2}
 \psi^{-},\qquad (\psi^{+})^{2}(\psi^{-})^{3}(\psi^{+})^{3}
 =\psi^{+}(\psi^{-})^{2}(\psi^{+})^{3}.
\label{eq:3alg6'}
\end{align}
Next, multiplying Eq.~(\ref{eq:3alg1}) by $(\psi^{+})^{3}\psi^{-}$ from
left or by $\psi^{+}(\psi^{-})^{3}$ from right, and applying
(\ref{eq:mult2}) and (\ref{eq:3alg0}), we have
\begin{subequations}
\label{eq:3alg7}
\begin{align}
(\psi^{+})^{3}(\psi^{-})^{2}\psi^{+}+(\psi^{+})^{3}(\psi^{-})^{3}
 (\psi^{+})^{2}=2(\psi^{+})^{3}\psi^{-},\\
\psi^{-}(\psi^{+})^{2}(\psi^{-})^{3}+(\psi^{-})^{2}(\psi^{+})^{3}
 (\psi^{-})^{3}=2\psi^{+}(\psi^{-})^{3}.
\end{align}
\end{subequations}
From Eqs.~(\ref{eq:3alg6}) and (\ref{eq:3alg7}), we obtain
\begin{subequations}
\label{eq:3alg8}
\begin{align}
(\psi^{+})^{3}(\psi^{-})^{2}\psi^{+}=(\psi^{+})^{3}(\psi^{-})^{3}
 (\psi^{+})^{2}=(\psi^{+})^{3}\psi^{-},\\
\psi^{-}(\psi^{+})^{2}(\psi^{-})^{3}=(\psi^{-})^{2}(\psi^{+})^{3}
 (\psi^{-})^{3}=\psi^{+}(\psi^{-})^{3},
\end{align}
\end{subequations}
and thus their sign-interchanged relations
\begin{subequations}
\label{eq:3alg8'}
\begin{align}
(\psi^{-})^{3}(\psi^{+})^{2}\psi^{-}=(\psi^{-})^{3}(\psi^{+})^{3}
 (\psi^{-})^{2}=(\psi^{-})^{3}\psi^{+},\\
\psi^{+}(\psi^{-})^{2}(\psi^{+})^{3}=(\psi^{+})^{2}(\psi^{-})^{3}
 (\psi^{+})^{3}=\psi^{-}(\psi^{+})^{3}.
\end{align}
\end{subequations}
Next, multiplying (\ref{eq:3alg1}) by two $\psi^{-}$s as $(\psi^{-})^{2}
\times$(\ref{eq:3alg1}), $\psi^{-}\times$(\ref{eq:3alg1})$\times
\psi^{-}$, and (\ref{eq:3alg1})$\times(\psi^{-})^{2}$, and applying
(\ref{eq:mult2}), (\ref{eq:3alg0}), and the formulas
(\ref{eq:3alg8})--(\ref{eq:3alg8'}), we get
\begin{align}
\{\psi^{+},(\psi^{-})^{3}\}&=(\psi^{-})^{2}(\psi^{+})^{2}
 (\psi^{-})^{2}=(\psi^{-})^{2},
\label{eq:3alg9}\\
\{\psi^{-},(\psi^{+})^{3}\}&=(\psi^{+})^{2}(\psi^{-})^{2}
 (\psi^{+})^{2}=(\psi^{+})^{2}.
\label{eq:3alg9'}
\end{align}
As a consequence of the formulas (\ref{eq:3alg9})--(\ref{eq:3alg9'}),
we immediately have
\begin{align}
\psi^{+}(\psi^{-})^{3}\psi^{+}&=\psi^{+}(\psi^{-})^{2}-(\psi^{+})^{2}
 (\psi^{-})^{3}=(\psi^{-})^{2}\psi^{+}-(\psi^{-})^{3}(\psi^{+})^{2},
\label{eq:3alg10}\\
\psi^{-}(\psi^{+})^{3}\psi^{-}&=\psi^{-}(\psi^{+})^{2}-(\psi^{-})^{2}
 (\psi^{+})^{3}=(\psi^{+})^{2}\psi^{-}-(\psi^{+})^{3}(\psi^{-})^{2},
\label{eq:3alg10'}
\end{align}
and as by products
\begin{align}
[\psi^{+},(\psi^{-})^{2}]=[(\psi^{+})^{2},(\psi^{-})^{3}],\qquad
 [\psi^{-},(\psi^{+})^{2}]=[(\psi^{-})^{2},(\psi^{+})^{3}].
\label{eq:3alg11}
\end{align}
From Eqs.~(\ref{eq:3alg3}) and (\ref{eq:3alg9}), we have
\begin{align}
(\psi^{+})^{2}(\psi^{-})^{3}\psi^{+}=\psi^{-}(\psi^{+})^{3}
 (\psi^{-})^{2}&=(\psi^{+})^{2}(\psi^{-})^{2}-(\psi^{+})^{3}
 (\psi^{-})^{3},
\label{eq:3alg12}\\
(\psi^{-})^{2}(\psi^{+})^{3}\psi^{-}=\psi^{+}(\psi^{-})^{3}
 (\psi^{+})^{2}&=(\psi^{-})^{2}(\psi^{+})^{2}-(\psi^{-})^{3}
 (\psi^{+})^{3}.
\label{eq:3alg12'}
\end{align}
Next, we examine the combination $(\psi^{-})^{2}(\psi^{+})^{3}
(\psi^{-})^{2}$. From the associativity and the formulas
(\ref{eq:3alg8'}), (\ref{eq:3alg10}), and (\ref{eq:3alg12'}) we
have on one hand
\begin{align}
(\psi^{-})^{2}(\psi^{+})^{3}(\psi^{-})^{2}&=\left((\psi^{-})^{2}
 (\psi^{+})^{3}\psi^{-}\right)\psi^{-}=\left(\psi^{+}(\psi^{-})^{3}
 (\psi^{+})^{2}\right)\psi^{-}\notag\\
&=\psi^{+}\left((\psi^{-})^{3}(\psi^{+})^{2}\psi^{-}\right)
 =\psi^{+}(\psi^{-})^{3}\psi^{+}\notag\\
&=(\psi^{-})^{2}\psi^{+}-(\psi^{-})^{3}(\psi^{+})^{2}.
\label{eq:3alg13}
\end{align}
On the other hand, from the formulas (\ref{eq:3alg5'}) and
(\ref{eq:3alg12'}) we have
\begin{align}
(\psi^{-})^{2}(\psi^{+})^{3}(\psi^{-})^{2}&=\left((\psi^{-})^{2}
 (\psi^{+})^{3}\psi^{-}\right)\psi^{-}=\left((\psi^{-})^{2}
(\psi^{+})^{2}-(\psi^{-})^{3}(\psi^{+})^{3}\right)\psi^{-}\notag\\
&=(\psi^{-})^{2}(\psi^{+})^{2}\psi^{-}-(\psi^{-})^{3}(\psi^{+})^{2}.
\end{align}
In a similar way, we further obtain for the same combination
\begin{align}
(\psi^{-})^{2}(\psi^{+})^{3}(\psi^{-})^{2}&=\psi^{+}(\psi^{-})^{2}
 -(\psi^{+})^{2}(\psi^{-})^{3}\notag\\
&=\psi^{-}(\psi^{+})^{2}(\psi^{-})^{2}-(\psi^{+})^{2}(\psi^{-})^{3}.
\label{eq:3alg14}
\end{align}
Comparing Eqs.~(\ref{eq:3alg13})--(\ref{eq:3alg14}), we finally obtain
\begin{align}
(\psi^{-})^{2}(\psi^{+})^{3}(\psi^{-})^{2}=\psi^{+}(\psi^{-})^{2}
 -(\psi^{+})^{2}(\psi^{-})^{3}=(\psi^{-})^{2}\psi^{+}-(\psi^{-})^{3}
 (\psi^{+})^{2},
\label{eq:3alg15}\\
(\psi^{-})^{2}(\psi^{+})^{2}\psi^{-}=(\psi^{-})^{2}\psi^{+},\qquad
 \psi^{-}(\psi^{+})^{2}(\psi^{-})^{2}=\psi^{+}(\psi^{-})^{2},
\label{eq:3alg16}
\end{align}
and their sign-interchanged relations
\begin{align}
(\psi^{+})^{2}(\psi^{-})^{3}(\psi^{+})^{2}=\psi^{-}(\psi^{+})^{2}
 -(\psi^{-})^{2}(\psi^{+})^{3}=(\psi^{+})^{2}\psi^{-}-(\psi^{+})^{3}
 (\psi^{-})^{2},
\label{eq:3alg15'}\\
(\psi^{+})^{2}(\psi^{-})^{2}\psi^{+}=(\psi^{+})^{2}\psi^{-},\qquad
 \psi^{+}(\psi^{-})^{2}(\psi^{+})^{2}=\psi^{-}(\psi^{+})^{2}.
\label{eq:3alg16'}
\end{align}
For the combination $(\psi^{-})^{2}(\psi^{+})^{3}\psi^{-}$, we have
with the aid of Eqs.~(\ref{eq:3alg10'}) and (\ref{eq:3alg12})
\begin{align}
(\psi^{-})^{2}(\psi^{+})^{3}\psi^{-}&=\psi^{-}\left(\psi^{-}
 (\psi^{+})^{3}\psi^{-}\right)=\psi^{-}\left((\psi^{+})^{2}\psi^{-}
 -(\psi^{+})^{3}(\psi^{-})^{2}\right)\notag\\
&=\psi^{-}(\psi^{+})^{2}\psi^{-}-(\psi^{+})^{2}(\psi^{-})^{2}
 +(\psi^{+})^{3}(\psi^{-})^{3}.
\label{eq:3alg17}
\end{align}
Comparing Eqs.~(\ref{eq:3alg12'}) and (\ref{eq:3alg17}), we obtain
\begin{align}
\psi^{-}(\psi^{+})^{2}\psi^{-}=\psi^{+}(\psi^{-})^{2}\psi^{+}
 =\{(\psi^{-})^{2},(\psi^{+})^{2}\}-\{(\psi^{-})^{3},(\psi^{+})^{3}\}.
\label{eq:3alg18}
\end{align}
Finally, multiplying (\ref{eq:3alg1}) by $\psi^{-}$ from left or right,
and applying (\ref{eq:mult2}), (\ref{eq:3alg0}), (\ref{eq:3alg5}),
(\ref{eq:3alg5'}), and (\ref{eq:3alg16}), we get
\begin{align}
\{\psi^{+},(\psi^{-})^{2}\}+\{(\psi^{+})^{2},(\psi^{-})^{3}\}
 &=2\psi^{-},
\label{eq:3alg19}\\
\{\psi^{-},(\psi^{+})^{2}\}+\{(\psi^{-})^{2},(\psi^{+})^{3}\}
 &=2\psi^{+}.
\label{eq:3alg20}
\end{align}
Multiplying the formula (\ref{eq:3alg19}) by $\psi^{+}$ from left,
and applying Eqs.~(\ref{eq:3alg12'}) and (\ref{eq:3alg18}), we further
obtain
\begin{align}
2\{(\psi^{-})^{2},(\psi^{+})^{2}\}&=2\left(\psi^{+}\psi^{-}+(\psi^{-})^{3}
 (\psi^{+})^{3}\right)=2\left(\psi^{-}\psi^{+}+(\psi^{+})^{3}(\psi^{-})^{3}
 \right)\notag\\
&=\{\psi^{-},\psi^{+}\}+\{(\psi^{-})^{3},(\psi^{+})^{3}\}.
\label{eq:3alg21}
\end{align}
Hence, the relation (\ref{eq:3alg1}) is decomposed as
\begin{align}
\{\psi^{-},\psi^{+}\}+\{(\psi^{-})^{3},(\psi^{+})^{3}\}=2I,\qquad
 \{(\psi^{-})^{2},(\psi^{+})^{2}\}=I.
\label{eq:3alg22}
\end{align}
The second equality in Eq.~(\ref{eq:3alg21}) can be expressed as
\begin{align}
[\psi^{-},\psi^{+}]=[(\psi^{-})^{3},(\psi^{+})^{3}].
\end{align}
Summarizing the results, we have derived third-order
parafermionic algebra as follows:
\begin{align}
(\psi^{-})^{4}=(\psi^{+})^{4}=0,
\label{eq:alg3a}\\
\{\psi^{-},(\psi^{+})^{3}\}=(\psi^{+})^{2},\qquad
 \{\psi^{+},(\psi^{-})^{3}\}=(\psi^{-})^{2},
\label{eq:alg3b}\\
\{\psi^{-},(\psi^{+})^{2}\}+\{(\psi^{-})^{2},(\psi^{+})^{3}\}
 =2\psi^{+},\qquad\{\psi^{+},(\psi^{-})^{2}\}+\{(\psi^{+})^{2},
 (\psi^{-})^{3}\}=2\psi^{-},
\label{eq:alg3c}\\
\{\psi^{-},\psi^{+}\}+\{(\psi^{-})^{3},(\psi^{+})^{3}\}=2I,
 \qquad\{(\psi^{-})^{2},(\psi^{+})^{2}\}=I,
\label{eq:alg3d}\\
[\psi^{-},(\psi^{+})^{2}]=[(\psi^{-})^{2},(\psi^{+})^{3}],\qquad
 [\psi^{+},(\psi^{-})^{2}]=[(\psi^{+})^{2},(\psi^{-})^{3}],
\label{eq:alg3e}\\
[\psi^{-},\psi^{+}]=[(\psi^{-})^{3},(\psi^{+})^{3}].
\label{eq:alg3f}
\end{align}
The relations (\ref{eq:alg3c}) and (\ref{eq:alg3e})
together imply important formulas
\begin{align}
(\psi^{-})^{2}\psi^{+}+(\psi^{+})^{2}(\psi^{-})^{3}&=
 \psi^{+}(\psi^{-})^{2}+(\psi^{-})^{3}(\psi^{+})^{2}=\psi^{-},
\label{eq:alg3g}\\
\psi^{-}(\psi^{+})^{2}+(\psi^{+})^{3}(\psi^{-})^{2}&=
 (\psi^{+})^{2}\psi^{-}+(\psi^{-})^{2}(\psi^{+})^{3}=\psi^{+}.
\label{eq:alg3h}
\end{align}
Similarly, the relations (\ref{eq:alg3d}) and (\ref{eq:alg3f}) together
imply
\begin{align}
\psi^{-}\psi^{+}+(\psi^{+})^{3}(\psi^{-})^{3}
 =\psi^{+}\psi^{-}+(\psi^{-})^{3}(\psi^{+})^{3}=I.
\label{eq:alg3i}
\end{align}
Classifying the formulas (\ref{eq:3alg2})--(\ref{eq:3alg2'}),
(\ref{eq:3alg5})--(\ref{eq:3alg6'}), (\ref{eq:3alg8})--(\ref{eq:3alg10'}),
(\ref{eq:3alg12})--(\ref{eq:3alg12'}), (\ref{eq:3alg15})--%
(\ref{eq:3alg16'}), and (\ref{eq:3alg18}) with respect to
parafermionic degrees, we have the following multiplication
law for the parafermionic operators of order 3:
\begin{itemize}
\item Degree 0:
 \begin{align*}
 (\psi^{-})^{3}\psi^{+}(\psi^{-})^{2}&=0,
  & \psi^{+}(\psi^{-})^{3}(\psi^{+})^{2}&=\Pi_{1},\\
 (\psi^{-})^{3}(\psi^{+})^{2}(\psi^{-})^{3}&=0,
  & \psi^{+}(\psi^{-})^{2}\psi^{+}&=\Pi_{1}+\Pi_{2},\\
 (\psi^{-})^{2}\psi^{+}(\psi^{-})^{3}&=0,
  & (\psi^{+})^{2}(\psi^{-})^{3}\psi^{+}&=\Pi_{2},\\
 (\psi^{-})^{2}(\psi^{+})^{3}\psi^{-}&=\Pi_{1},
  & (\psi^{+})^{2}\psi^{-}(\psi^{+})^{3}&=0,\\
 \psi^{-}(\psi^{+})^{2}\psi^{-}&=\Pi_{1}+\Pi_{2},
  & (\psi^{+})^{3}(\psi^{-})^{2}(\psi^{+})^{3}&=0,\\
 \psi^{-}(\psi^{+})^{3}(\psi^{-})^{2}&=\Pi_{2},
  & (\psi^{+})^{3}\psi^{-}(\psi^{+})^{2}&=0.
 \end{align*}
\item Degree 1:
 \begin{align*}
 (\psi^{-})^{3}\psi^{+}\psi^{-}&=(\psi^{-})^{3},
  & \psi^{+}(\psi^{-})^{3}(\psi^{+})^{3}&=(\psi^{-})^{2}
  (\psi^{+})^{3},\\
 (\psi^{-})^{3}(\psi^{+})^{3}(\psi^{-})^{3}&=(\psi^{-})^{3},
  & \psi^{+}\psi^{-}\psi^{+}&=\psi^{+},\\
 \psi^{-}\psi^{+}(\psi^{-})^{3}&=(\psi^{-})^{3},
  & (\psi^{+})^{3}(\psi^{-})^{3}\psi^{+}&=(\psi^{+})^{3}
  (\psi^{-})^{2},\\
 (\psi^{-})^{3}(\psi^{+})^{2}(\psi^{-})^{2}&=(\psi^{-})^{3},
  & \psi^{+}(\psi^{-})^{2}(\psi^{+})^{2}&=\psi^{-}(\psi^{+})^{2},\\
 (\psi^{-})^{2}\psi^{+}(\psi^{-})^{2}&=(\psi^{-})^{3},
  & (\psi^{+})^{2}(\psi^{-})^{3}(\psi^{+})^{2}&=\varphi^{+},\\
 (\psi^{-})^{2}(\psi^{+})^{2}(\psi^{-})^{3}&=(\psi^{-})^{3},
  & (\psi^{+})^{2}(\psi^{-})^{2}\psi^{+}&=(\psi^{+})^{2}\psi^{-},\\
 \psi^{-}(\psi^{+})^{3}\psi^{-}&=\varphi^{+},
  & (\psi^{+})^{3}\psi^{-}(\psi^{+})^{3}&=0.
 \end{align*}
\item Degree 2:
 \begin{align*}
 (\psi^{-})^{3}(\psi^{+})^{2}\psi^{-}&=(\psi^{-})^{3}\psi^{+},
  & \psi^{+}(\psi^{-})^{2}(\psi^{+})^{3}&=\psi^{-}(\psi^{+})^{3},\\
 (\psi^{-})^{3}(\psi^{+})^{3}(\psi^{-})^{2}&=(\psi^{-})^{3}\psi^{+},
  & \psi^{+}\psi^{-}(\psi^{+})^{2}&=(\psi^{+})^{2},\\
 (\psi^{-})^{2}\psi^{+}\psi^{-}&=(\psi^{-})^{2},
  & (\psi^{+})^{2}(\psi^{-})^{3}(\psi^{+})^{3}&=\psi^{-}
  (\psi^{+})^{3},\\
 (\psi^{-})^{2}(\psi^{+})^{3}(\psi^{-})^{3}&=\psi^{+}(\psi^{-})^{3},
  & (\psi^{+})^{2}\psi^{-}\psi^{+}&=(\psi^{+})^{2},\\
 \psi^{-}\psi^{+}(\psi^{-})^{2}&=(\psi^{-})^{2},
  & (\psi^{+})^{3}(\psi^{-})^{3}(\psi^{+})^{2}&=(\psi^{+})^{3}
  \psi^{-},\\
 \psi^{-}(\psi^{+})^{2}(\psi^{-})^{3}&=\psi^{+}(\psi^{-})^{3},
  & (\psi^{+})^{3}(\psi^{-})^{2}\psi^{+}&=(\psi^{+})^{3}\psi^{-},\\
 (\psi^{-})^{2}(\psi^{+})^{2}(\psi^{-})^{2}&=(\psi^{-})^{2},
  & (\psi^{+})^{2}(\psi^{-})^{2}(\psi^{+})^{2}&=(\psi^{+})^{2}.
 \end{align*}
\item Degree 3:
 \begin{align*}
 (\psi^{-})^{3}\psi^{+}(\psi^{-})^{3}&=0,
  & \psi^{+}(\psi^{-})^{3}\psi^{+}&=\varphi^{-},\\
 (\psi^{-})^{3}(\psi^{+})^{3}\psi^{-}&=(\psi^{-})^{3}(\psi^{+})^{2},
  & \psi^{+}\psi^{-}(\psi^{+})^{3}&=(\psi^{+})^{3},\\
 \psi^{-}\psi^{+}\psi^{-}&=\psi^{-},
  & (\psi^{+})^{3}(\psi^{-})^{3}(\psi^{+})^{3}&=(\psi^{+})^{3},\\
 \psi^{-}(\psi^{+})^{3}(\psi^{-})^{3}&=(\psi^{+})^{2}(\psi^{-})^{3},
  & (\psi^{+})^{3}\psi^{-}\psi^{+}&=(\psi^{+})^{3},\\
 (\psi^{-})^{2}(\psi^{+})^{2}\psi^{-}&=(\psi^{-})^{2}\psi^{+},
  & (\psi^{+})^{2}(\psi^{-})^{2}(\psi^{+})^{3}&=(\psi^{+})^{3},\\
 (\psi^{-})^{2}(\psi^{+})^{3}(\psi^{-})^{2}&=\varphi^{-},
  & (\psi^{+})^{2}\psi^{-}(\psi^{+})^{2}&=(\psi^{+})^{3},\\
 \psi^{-}(\psi^{+})^{2}(\psi^{-})^{2}&=\psi^{+}(\psi^{-})^{2},
  & (\psi^{+})^{3}(\psi^{-})^{2}(\psi^{+})^{2}&=(\psi^{+})^{3}.
 \end{align*}
\end{itemize}
In the above, $\varphi^{-}$, $\varphi^{+}$, $\Pi_{1}$, and $\Pi_{2}$
are defined by
\begin{align}
\varphi^{-}&=\psi^{+}(\psi^{-})^{2}-(\psi^{+})^{2}(\psi^{-})^{3}
 =(\psi^{-})^{2}\psi^{+}-(\psi^{-})^{3}(\psi^{+})^{2},\\
\varphi^{+}&=\psi^{-}(\psi^{+})^{2}-(\psi^{-})^{2}(\psi^{+})^{3}
 =(\psi^{+})^{2}\psi^{-}-(\psi^{+})^{3}(\psi^{-})^{2},\\
\Pi_{1}&=\psi^{+}\psi^{-}-(\psi^{+})^{2}(\psi^{-})^{2}
 =(\psi^{-})^{2}(\psi^{+})^{2}-(\psi^{-})^{3}(\psi^{+})^{3},
\label{eq:3pjo1}\\
\Pi_{2}&=(\psi^{+})^{2}(\psi^{-})^{2}-(\psi^{+})^{3}(\psi^{-})^{3}
 =\psi^{-}\psi^{+}-(\psi^{-})^{2}(\psi^{+})^{2}.
\end{align}
The above $\Pi_{1}$ and $\Pi_{2}$ are indeed the projection
operators into the 1- and 2-parafermionic subspaces, respectively.
The other projection operators $\Pi_{0}$ and $\Pi_{3}$ are given by
\begin{align}
\Pi_{0}=(\psi^{-})^{3}(\psi^{+})^{3},\qquad
 \Pi_{3}=(\psi^{+})^{3}(\psi^{-})^{3}.
\label{eq:3pjo3}
\end{align}
We can easily check with the aid of the multiplication law and
the formula (\ref{eq:alg3i}) that the operators $\Pi_{i}$ ($i=0,1,2,3$)
in Eqs.~(\ref{eq:3pjo1})--(\ref{eq:3pjo3}) satisfy
the definition (\ref{eq:defpo}) for $p=3$.
The intertwining relations  (\ref{eq:Pipsi}) can be easily checked as
\begin{align*}
\Pi_{1}\psi^{+}=\psi^{+}\Pi_{0}=(\psi^{-})^{2}(\psi^{+})^{3},\quad
 \Pi_{2}\psi^{+}=\psi^{+}\Pi_{1}=\varphi^{+},\quad
 \Pi_{3}\psi^{+}=\psi^{+}\Pi_{2}=(\psi^{+})^{3}(\psi^{-})^{2},\\
\psi^{-}\Pi_{1}=\Pi_{0}\psi^{-}=(\psi^{-})^{3}(\psi^{+})^{2},\quad
 \psi^{-}\Pi_{2}=\Pi_{1}\psi^{-}=\varphi^{-},\quad
 \psi^{-}\Pi_{3}=\Pi_{2}\psi^{-}=(\psi^{+})^{2}(\psi^{-})^{3}.
\end{align*}
We also note that the third-order parasuperalgebra
(\ref{eq:alg3a})--(\ref{eq:alg3i}) is consistent with
Eq.~(\ref{eq:defpfs}). From the relations (\ref{eq:alg3b}),
(\ref{eq:alg3h}), and (\ref{eq:alg3i}) we have
\begin{align*}
\psi^{-}|1\rangle&=\psi^{-}\psi^{+}|0\rangle
 =(I-(\psi^{+})^{3}(\psi^{-})^{3})|0\rangle=|0\rangle,\\
\psi^{-}|2\rangle&=\psi^{-}(\psi^{+})^{2}|0\rangle
 =(\psi^{+}-(\psi^{+})^{3}(\psi^{-})^{2})|0\rangle=|1\rangle,\\
\psi^{-}|3\rangle&=\psi^{-}(\psi^{+})^{3}|0\rangle
 =((\psi^{+})^{2}-(\psi^{+})^{3}\psi^{-})|0\rangle=|2\rangle,
\end{align*}
which are exactly Eq.~(\ref{eq:defpfs}). The assumption
(\ref{eq:mult3}) for $p=3$ is also satisfied as
\begin{align*}
(\psi^{-})^{3}\psi^{+}\Pi_{2}=(\psi^{-})^{2}\Pi_{2}=(\psi^{-})^{3}
 \psi^{+},\qquad\psi^{+}(\psi^{-})^{3}\Pi_{3}=(\psi^{-})^{2}\Pi_{3}
 =\psi^{+}(\psi^{-})^{3}.
\end{align*}
Therefore, we have confirmed that all the postulates in
Section~\ref{sec:rev} are fulfilled. In contrast to the second-order,
the trilinear relations of parafermionic statistics \cite{Gr52,OK82}
are not satisfied since
\begin{align*}
[\psi^{\mp},[\psi^{\pm},\psi^{\mp}]]=2\psi^{\mp}\psi^{\pm}\psi^{\mp}
 -\{\psi^{\pm},(\psi^{\mp})^{2}\}=2\psi^{\mp}-\{\psi^{\pm},
(\psi^{\mp})^{2}\},
\end{align*}
and the anti-commutators $\{\psi^{\pm},(\psi^{\mp})^{2}\}$ themselves
do not any more reduce to simpler forms in the third-order. However, we
find that the following quadrilinear relations are instead satisfied:
\begin{align}
\{\psi^{\mp},[\psi^{\mp},[\psi^{\pm},\psi^{\mp}]]\}
 =-[\psi^{\mp},[\psi^{\mp},\{\psi^{\pm},\psi^{\mp}\}]]
 =[\psi^{\mp},\{\psi^{\mp},[\psi^{\pm},\psi^{\mp}]\}]
 =(\psi^{\mp})^{2}.
\label{eq:quadri}
\end{align}
It is evident that the first quadrilinear term in Eq.~(\ref{eq:quadri})
is proportional to $(\psi^{\mp})^{2}$ if the trilinear relations of
parafermionic statistics $[\psi^{\mp},[\psi^{\pm},\psi^{\mp}]]\propto
\psi^{\mp}$ hold. Hence, the quadrilinear relations (\ref{eq:quadri})
can be regarded as generalized parafermionic statistics.

It is worth noticing that if we restrict the fundamental elements to
the set $\{I,(\psi^{-})^{2},(\psi^{+})^{2}\}$, the third-order
parafermionic algebra and multiplication law only consist of
\begin{align}
\{(\psi^{-})^{2},(\psi^{-})^{2}\}=\{(\psi^{+})^{2},(\psi^{+})^{2}\}
 =0,\qquad\{(\psi^{-})^{2},(\psi^{+})^{2}\}=I,\\
(\psi^{-})^{2}(\psi^{+})^{2}(\psi^{-})^{2}=(\psi^{-})^{2},\qquad
 (\psi^{+})^{2}(\psi^{-})^{2}(\psi^{+})^{2}=(\psi^{+})^{2},
\end{align}
which are exactly equivalent to the ordinary fermionic relations.
Therefore, in the third-order parafermionic system, $(\psi^{-})^{2}$
and $(\psi^{+})^{2}$ behave as ordinary fermions.

\section{Third-Order Parasupersymmetric Quantum Systems}
\label{sec:3psqs}

We are now in a position to construct third-order parasupersymmetric
quantum systems by using the third-order parafermionic algebra
just derived in the previous section. From Eqs.~(\ref{eq:3pjo1})--%
(\ref{eq:3pjo3}), and the multiplication law, the triple $(\bH,
\bQ^{-},\bQ^{+})$ in Eqs.~(\ref{eq:pfqs}) and (\ref{eq:pfsc}) for
$p=3$ is expressed as
\begin{align}
\bH=&\, H_{0}(\psi^{-})^{3}(\psi^{+})^{3}
 +H_{1}(\psi^{+}\psi^{-}-(\psi^{+})^{2}(\psi^{-})^{2})\notag\\
 &\,+H_{2}((\psi^{+})^{2}(\psi^{-})^{2}-(\psi^{+})^{3}(\psi^{-})^{3})
 +H_{3}(\psi^{+})^{3}(\psi^{-})^{3},\\
\bQ^{-}=&\, Q_{1}^{-}(\psi^{-})^{3}(\psi^{+})^{2}+Q_{2}^{-}\varphi^{-}
 +Q_{3}^{-}(\psi^{+})^{2}(\psi^{-})^{3},
\label{eq:psc3-}\\
\bQ^{+}=&\, Q_{1}^{+}(\psi^{-})^{2}(\psi^{+})^{3}+Q_{2}^{+}\varphi^{+}
 +Q_{3}^{+}(\psi^{+})^{3}(\psi^{-})^{2}.
\label{eq:psc3+}
\end{align}
We recall the fact that the above third-order parasupercharges
(\ref{eq:psc3-}) and (\ref{eq:psc3+}) already satisfy
the nilpotent condition (\ref{eq:pfsc1}) for $p=3$,
$(\bQ^{-})^{4}=(\bQ^{+})^{4}=0$, as we have previously mentioned in
Section~\ref{sec:rev}.

Now that we have had the third-order parafermionic algebra and
multiplication law, we can explicitly construct the `discrete-type'
conserved charges.
In the case of third-order, we can have twelve additional charges
$\bQ_{\{n\}}^{\pm}$ and $\bQ_{[n]}^{\pm}$ ($n=1,2,3$) defined
by Eq.~(\ref{eq:discc}). Among them, we find that there are
essentially four linearly independent new conserved charges, e.g.,
\begin{subequations}
\label{eq:3dch1}
\begin{align}
\bQ_{[2]}^{-}&=2Q_{2}^{-}\varphi^{-},&
 \bQ_{\{3\}}^{-}&=Q_{1}^{-}(\psi^{-})^{3}(\psi^{+})^{2}
 -Q_{3}^{-}(\psi^{+})^{2}(\psi^{-})^{3},\\
\bQ_{[2]}^{+}&=-2Q_{2}^{+}\varphi^{+},&
 \bQ_{\{3\}}^{+}&=-Q_{1}^{+}(\psi^{-})^{2}(\psi^{+})^{3}
 +Q_{3}^{+}(\psi^{+})^{3}(\psi^{-})^{2},
\end{align}
\end{subequations}
and the others are expressed as
\begin{subequations}
\begin{align}
\bQ_{\{1\}}^{-}=-\bQ_{\{3\}}^{-},\quad\bQ_{\{2\}}^{-}=0,\quad
 2\bQ_{[1]}^{-}=2\bQ_{[3]}^{-}=2\bQ^{-}-\bQ_{[2]}^{-},\\
2\bQ_{\{1\}}^{+}=-2\bQ_{\{3\}}^{+}+\bQ_{[2]}^{+},\quad
 \bQ_{\{2\}}^{+}=0,\\
4\bQ_{[1]}^{+}=-2\bQ^{+}-\bQ_{[2]}^{+}-2\bQ_{\{3\}}^{+},\quad
 2\bQ_{[3]}^{+}=-2\bQ^{+}-\bQ_{[2]}^{+}.
\end{align}
\end{subequations}
Needless to say, any linear combination of the original and new
parasupercharges, $\bQ^{\pm}$, $\bQ_{[n]}^{\pm}$, and $\bQ_{\{n\}}^{\pm}$,
is also conserved. In particular, the following combinations
\begin{subequations}
\begin{align}
\bQ_{2}^{-}&\equiv\bQ^{-}-\bQ_{[2]}^{-}=Q_{1}^{-}(\psi^{-})^{3}
 (\psi^{+})^{2}-Q_{2}^{-}\varphi^{-}+Q_{3}^{-}(\psi^{+})^{2}
 (\psi^{-})^{3},\\
\bQ_{3}^{-}&\equiv\frac{1}{2}\bQ_{[2]}^{-}+\bQ_{\{3\}}^{-}=Q_{1}^{-}
 (\psi^{-})^{3}(\psi^{+})^{2}+Q_{2}^{-}\varphi^{-}-Q_{3}^{-}
 (\psi^{+})^{2}(\psi^{-})^{3},
\end{align}
\end{subequations}
and their `adjoint' ones are exactly the conserved charges (of order $3$)
reported in Refs.~\cite{Kh92,Kh93}, which are generalizations of
the second-order ones in Ref.~\cite{DV90a}.

From Eqs.~(\ref{eq:inter}) and (\ref{eq:parac}), the
commutativity (\ref{eq:pfsc2}) and the non-linear constraints
(\ref{eq:pfsc3}) for $p=3$
\begin{subequations}
\begin{align}
(\bQ^{-})^{3}\bQ^{+}+(\bQ^{-})^{2}\bQ^{+}\bQ^{-}+\bQ^{-}\bQ^{+}
 (\bQ^{-})^{2}+\bQ^{+}(\bQ^{-})^{3}=C_{3}(\bQ^{-})^{2}\bH,\\
(\bQ^{+})^{3}\bQ^{-}+(\bQ^{+})^{2}\bQ^{-}\bQ^{+}+\bQ^{+}\bQ^{-}
 (\bQ^{+})^{2}+\bQ^{-}(\bQ^{+})^{3}=C_{3}\bH(\bQ^{+})^{2},
\end{align}
\end{subequations}
hold if and only if the following conditions
\begin{align}
H_{0}Q_{1}^{-}=Q_{1}^{-}H_{1},\quad
 H_{1}Q_{2}^{-}=Q_{2}^{-}H_{2},\quad
 H_{2}Q_{3}^{-}=Q_{3}^{-}H_{3},
\label{eq:3psc1}\\
Q_{1}^{-}Q_{2}^{-}Q_{3}^{-}Q_{3}^{+}+Q_{1}^{-}Q_{2}^{-}Q_{2}^{+}
 Q_{2}^{-}+Q_{1}^{-}Q_{1}^{+}Q_{1}^{-}Q_{2}^{-}
 =C_{3}Q_{1}^{-}Q_{2}^{-}H_{2},
\label{eq:3psc2}\\
Q_{2}^{-}Q_{3}^{-}Q_{3}^{+}Q_{3}^{-}+Q_{2}^{-}Q_{2}^{+}Q_{2}^{-}
 Q_{3}^{-}+Q_{1}^{+}Q_{1}^{-}Q_{2}^{-}Q_{3}^{-}
 =C_{3}Q_{2}^{-}Q_{3}^{-}H_{3},
\label{eq:3psc3}
\end{align}
and their `adjoint' relations
\begin{align}
Q_{1}^{+}H_{0}=H_{1}Q_{1}^{+},\quad
 Q_{2}^{+}H_{1}=H_{2}Q_{2}^{+},\quad
 Q_{3}^{+}H_{2}=H_{3}Q_{3}^{+},
\label{eq:3psc1'}\\
Q_{2}^{+}Q_{1}^{+}Q_{1}^{-}Q_{1}^{+}+Q_{2}^{+}Q_{2}^{-}Q_{2}^{+}
 Q_{1}^{+}+Q_{3}^{-}Q_{3}^{+}Q_{2}^{+}Q_{1}^{+}
 =C_{3}H_{2}Q_{2}^{+}Q_{1}^{+},
\label{eq:3psc2'}\\
Q_{3}^{+}Q_{2}^{+}Q_{1}^{+}Q_{1}^{-}+Q_{3}^{+}Q_{2}^{+}Q_{2}^{-}
 Q_{2}^{+}+Q_{3}^{+}Q_{3}^{-}Q_{3}^{+}Q_{2}^{+}
 =C_{3}H_{3}Q_{3}^{+}Q_{2}^{+},
\label{eq:3psc3'}
\end{align}
are satisfied. We note that when a solution to Eq.~(\ref{eq:3psc2})
and (\ref{eq:3psc2'}) are given by
\begin{align}
C_{3}Q_{2}^{-}H_{2}=Q_{1}^{+}Q_{1}^{-}Q_{2}^{-}+Q_{2}^{-}Q_{2}^{+}
 Q_{2}^{-}+Q_{2}^{-}Q_{3}^{-}Q_{3}^{+},
\label{eq:3psc4}
\end{align}
and its `adjoint' relation, the conditions (\ref{eq:3psc3}) and
(\ref{eq:3psc3'}) are automatically satisfied so long as the third
intertwining relations in Eqs.~(\ref{eq:3psc1}) and
(\ref{eq:3psc1'}) hold.
Thus, in this case it is sufficient to solve the intertwining
relations (\ref{eq:3psc1}) and (\ref{eq:3psc1'}).
In general, we do not need to solve the `adjoint' conditions.

For the third-order case, we have two different
quasi-parasupersymmetries, namely, those of order $(3,2)$ and
$(3,3)$. The conditions are given by
Eqs.~(\ref{eq:3psc1})--(\ref{eq:3psc3'}) but the first-order
intertwining relations (\ref{eq:3psc1}) and (\ref{eq:3psc1'}) are
replaced in the case of order $(3,2)$ by the second-order
intertwining relations
\begin{subequations}
\label{eq:3qpsc1}
\begin{align}
H_{0}Q_{1}^{-}Q_{2}^{-}=Q_{1}^{-}Q_{2}^{-}H_{2},\qquad
 H_{1}Q_{2}^{-}Q_{3}^{-}=Q_{2}^{-}Q_{3}^{-}H_{3},\\
Q_{2}^{+}Q_{1}^{+}H_{0}=H_{2}Q_{2}^{+}Q_{1}^{+},\qquad
 Q_{3}^{+}Q_{2}^{+}H_{1}=H_{3}Q_{3}^{+}Q_{2}^{+},
\end{align}
\end{subequations}
and in the case of order $(3,3)$ by the third-order ones
\begin{align}
H_{0}Q_{1}^{-}Q_{2}^{-}Q_{3}^{-}=Q_{1}^{-}Q_{2}^{-}Q_{3}^{-}
 H_{3},\qquad Q_{3}^{+}Q_{2}^{+}Q_{1}^{+}H_{0}=H_{3}
 Q_{3}^{+}Q_{2}^{+}Q_{1}^{+}.
\label{eq:3qpsc2}
\end{align}
In the followings, we will show two different representations
for the system $(H_{k},Q_{k}^{\pm})$ which satisfies
the parasupersymmetric conditions (\ref{eq:3psc1})--(\ref{eq:3psc3'})
or quasi-parasupersymmetric ones, (\ref{eq:3qpsc1}) or
(\ref{eq:3qpsc2}).

\subsection{One-Variable Representation}
\label{ssec:exam1}

First, we shall realize a third-order parasupersymmetric quantum
system of one degree of freedom. Let us put
\begin{align}
H_{k}=-\frac{1}{2}\frac{\rmd^{2}}{\rmd q^{2}}+V_{k}(q),
 \qquad Q_{k}^{\pm}=\pm\frac{\rmd}{\rmd q}+W_{k}(q).
\label{eq:1vrep1}
\end{align}
The general solutions to Eqs.~(\ref{eq:3psc1}) and (\ref{eq:3psc1'})
are given by
\begin{subequations}
\label{eq:1vrep2}
\begin{align}
2H_{0}=Q_{1}^{-}Q_{1}^{+}-2R_{1},\quad
 2H_{1}=Q_{1}^{+}Q_{1}^{-}-2R_{1}=Q_{2}^{-}Q_{2}^{+}-2R_{2},\\
2H_{2}=Q_{2}^{+}Q_{2}^{-}-2R_{2}=Q_{3}^{-}Q_{3}^{+}-2R_{3},
 \quad 2H_{3}=Q_{3}^{+}Q_{3}^{-}-2R_{3},
\end{align}
\end{subequations}
where $R_{i}$ ($i=1,2,3$) are constants and the functions $W_{k}$
($k=1,2,3$) must satisfy,
\begin{align}
W'_{k}+W'_{k+1}+W_{k}^{2}-W_{k+1}^{2}=2R_{k}-2R_{k+1}\quad(k=1,2).
\label{eq:1vrep3}
\end{align}
Substituting them into Eqs.~(\ref{eq:3psc2}) and (\ref{eq:3psc2'}),
we find that the conditions (\ref{eq:3psc2}) and (\ref{eq:3psc2'})
are satisfied if and only if
\begin{align}
C_{3}=6,\qquad R_{1}+R_{2}+R_{3}=0.
\label{eq:1vrep4}
\end{align}
In this case, the relation (\ref{eq:3psc4}) and its `adjoint' hold.
Hence, the remaining conditions (\ref{eq:3psc3}) and (\ref{eq:3psc3'})
are automatically satisfied. The third-order parasupersymmetric
quantum system (\ref{eq:1vrep1})--(\ref{eq:1vrep4}) is equivalent to
the generalized Rubakov--Spiridonov model of order $3$ constructed
in Ref.~\cite{Kh92,Kh93}.

For the second-order parasupersymmetric quantum system of
Rubakov--Spiridonov type, we found previously \cite{Ta07a} that
the system admits a generalized $2$-fold superalgebra. In what follows,
we show that the present third-order system analogously admits
a generalized $3$-fold superalgebra. To begin with, we construct
zeroth-degree operators composed of only $\bQ^{-}$ and $\bQ^{+}$.
Owing to the nilpotency $(\bQ^{\pm})^{4}=0$,
we can easily show that any such operator should be expressed as
a function of the following six composite operators:
\begin{align}
\bQ^{-}\bQ^{+}&=Q_{1}^{-}Q_{1}^{+}\Pi_{0}+Q_{2}^{-}Q_{2}^{+}\Pi_{1}
 +Q_{3}^{-}Q_{3}^{+}\Pi_{2},
\label{eq:mono1}\\
\bQ^{+}\bQ^{-}&=Q_{1}^{+}Q_{1}^{-}\Pi_{1}+Q_{2}^{+}Q_{2}^{-}\Pi_{2}
 +Q_{3}^{+}Q_{3}^{-}\Pi_{3},\\
(\bQ^{-})^{2}(\bQ^{+})^{2}&=Q_{1}^{-}Q_{2}^{-}Q_{2}^{+}Q_{1}^{+}\Pi_{0}
 +Q_{2}^{-}Q_{3}^{-}Q_{3}^{+}Q_{2}^{+}\Pi_{1},\\
(\bQ^{+})^{2}(\bQ^{-})^{2}&=Q_{2}^{+}Q_{1}^{+}Q_{1}^{-}Q_{2}^{-}\Pi_{2}
 +Q_{3}^{+}Q_{2}^{+}Q_{2}^{-}Q_{3}^{-}\Pi_{3},\\
(\bQ^{-})^{3}(\bQ^{+})^{3}&=Q_{1}^{-}Q_{2}^{-}Q_{3}^{-}
 Q_{3}^{+}Q_{2}^{+}Q_{1}^{+}\Pi_{0},\\
(\bQ^{+})^{3}(\bQ^{-})^{3}&=Q_{3}^{+}Q_{2}^{+}Q_{1}^{+}
 Q_{1}^{-}Q_{2}^{-}Q_{3}^{-}\Pi_{3}.
\label{eq:mono6}
\end{align}
Now, substituting the relations (\ref{eq:1vrep2}) into the above
(\ref{eq:mono1})--(\ref{eq:mono6}), we obtain
\begin{align}
\bQ^{-}\bQ^{+}&=2(H_{0}+R_{1})\Pi_{0}+2(H_{1}+R_{2})\Pi_{1}
 +2(H_{2}+R_{3})\Pi_{2},\\
\bQ^{+}\bQ^{-}&=2(H_{1}+R_{1})\Pi_{1}+2(H_{2}+R_{2})\Pi_{2}
 +2(H_{3}+R_{3})\Pi_{3},\\
(\bQ^{-})^{2}(\bQ^{+})^{2}&=4(H_{0}+R_{1})(H_{0}+R_{2})
 \Pi_{0}+4(H_{1}+R_{2})(H_{1}+R_{3})\Pi_{1},\\
(\bQ^{+})^{2}(\bQ^{-})^{2}&=4(H_{2}+R_{1})(H_{2}+R_{2})
 \Pi_{2}+4(H_{3}+R_{2})(H_{3}+R_{3})\Pi_{3},\\
(\bQ^{-})^{3}(\bQ^{+})^{3}&=8(H_{0}+R_{1})(H_{0}+R_{2})
 (H_{0}+R_{3})\Pi_{0},\\
(\bQ^{+})^{3}(\bQ^{-})^{3}&=8(H_{3}+R_{1})(H_{3}+R_{2})
 (H_{3}+R_{3})\Pi_{3}.
\end{align}
From these formulas, we can easily find non-linear relations as follows:
\begin{multline}
(\bQ^{-})^{3}(\bQ^{+})^{3}+\left\{
 \begin{array}{l}
 \bQ^{+}(\bQ^{-})^{3}(\bQ^{+})^{2}\\
 (\bQ^{-})^{2}(\bQ^{+})^{3}\bQ^{-}
 \end{array}\right\}+\left\{
 \begin{array}{l}
 \bQ^{-}(\bQ^{+})^{3}(\bQ^{-})^{2}\\
 (\bQ^{+})^{2}(\bQ^{-})^{3}\bQ^{+}
 \end{array}
 \right\}+(\bQ^{+})^{3}(\bQ^{-})^{3}\\
=8(\bH+R_{1})(\bH+R_{2})(\bH+R_{3}).
\label{eq:g3falg}
\end{multline}
As in the case of second-order, they can be regarded as generalizations
of $3$-fold superalgebra. Indeed, if we restrict the linear space
$\fF\times\sV_{3}$ on which the system $\bH$ acts to
$\fF\times(\sV_{3}^{(0)}\dotplus\sV_{3}^{(3)})$ (cf. the
definition between Eqs.~(\ref{eq:defpfs}) and (\ref{eq:defadv})),
we have
\begin{align}
\{(\bQ^{-})^{3},(\bQ^{+})^{3}\}=8(\bH+R_{1})(\bH+R_{2})(\bH+R_{3})
 \bigr|_{\fF\times(\sV_{3}^{(0)}\dotplus\sV_{3}^{(3)})}.
\end{align}
This, together with the trivial (anti-)commutation relations
\begin{align}
\{(\bQ^{-})^{3},(\bQ^{-})^{3}\}=\{(\bQ^{+})^{3},(\bQ^{+})^{3}\}
 =[(\bQ^{\pm})^{3},\bH]=0,
\end{align}
constitutes a type of $3$-fold superalgebra in the sector
$\fF\times(\sV_{3}^{(0)}\dotplus\sV_{3}^{(3)})$.\\

Next, we shall examine quasi-parasupersymmetry in the system given
by Eq.~(\ref{eq:1vrep1}). For the purpose, we first solve the conditions
(\ref{eq:3psc2}) and (\ref{eq:3psc2'}). Substituting the expression
(\ref{eq:1vrep1}) into the conditions (\ref{eq:3psc2}) or
(\ref{eq:3psc2'}), we can easily find that they are satisfied if and
only if $C_{3}=6$ and
\begin{align}
6V_{2}=W'_{1}+3W'_{2}-W'_{3}+W_{1}^{2}+W_{2}^{2}+W_{3}^{2},
\label{eq:qp1c1}\\
W''_{1}+W''_{2}+2W_{1}W'_{1}-2W_{2}W'_{2}=0.
\label{eq:qp1c2}
\end{align}
Similarly, if we substitute Eq.~(\ref{eq:1vrep1}) into the conditions
(\ref{eq:3psc3}) or (\ref{eq:3psc3'}) with $C_{3}=6$, we find
\begin{align}
6V_{3}=W'_{1}+3W'_{2}+5W'_{3}+W_{1}^{2}+W_{2}^{2}+W_{3}^{2},
\label{eq:qp1c3}\\
W''_{1}+2W''_{2}+W''_{3}+2W_{1}W'_{1}-2W_{3}W'_{3}=0,
\label{eq:qp1c4}
\end{align}
and
\begin{multline}
W'''_{1}+W'''_{2}+2W_{1}W''_{1}+2(W'_{1})^{2}-W''_{1}W_{2}
 +W''_{1}W_{3}-3W_{2}W''_{2}-2(W'_{2})^{2}\\
+W''_{2}W_{3}-2W_{1}W'_{1}W_{2}+2W_{1}W'_{1}W_{3}+2W_{2}^{2}W'_{2}
 -2W_{2}W'_{2}W_{3}=0.
\label{eq:qp1c5}
\end{multline}
It is easy to check that the condition (\ref{eq:qp1c2}) implies
(\ref{eq:qp1c5}). From Eqs.~(\ref{eq:qp1c2}) and (\ref{eq:qp1c4}),
we obtain exactly the same conditions as Eq.~(\ref{eq:1vrep3}).
Thus, from the potential forms (\ref{eq:qp1c1}) and (\ref{eq:qp1c3}),
and the condition (\ref{eq:1vrep3}) we have
\begin{subequations}
\label{eq:qp1c6}
\begin{align}
6H_{2}&=3Q_{2}^{+}Q_{2}^{-}+2(R_{1}-2R_{2}+R_{3})
 =3Q_{3}^{-}Q_{3}^{+}+2(R_{1}+R_{2}-2R_{3}),\\
6H_{3}&=3Q_{3}^{+}Q_{3}^{-}+2(R_{1}+R_{2}-2R_{3}).
\end{align}
\end{subequations}
We now first consider quasi-parasupersymmetry of order $(3,2)$. The
remaining conditions to be satisfied are Eqs.~(\ref{eq:3qpsc1}).
Substituting the expressions (\ref{eq:1vrep1}) and (\ref{eq:qp1c6}) into
the conditions (\ref{eq:3qpsc1}), we find that they are satisfied if
and only if
\begin{subequations}
\label{eq:qp1c7}
\begin{align}
2V_{0}&=W_{2}^{2}-2W'_{1}-W'_{2}+\frac{2}{3}(R_{1}-2R_{2}+R_{3}),\\
2V_{1}&=W_{3}^{2}-2W'_{2}-W'_{3}+\frac{2}{3}(R_{1}+R_{2}-2R_{3}),
\end{align}
\end{subequations}
and Eq.~(\ref{eq:1vrep3}) hold. Combining all the results
(\ref{eq:1vrep3}), (\ref{eq:qp1c6}), and (\ref{eq:qp1c7}), we finally
derive the necessary and sufficient condition for the system
(\ref{eq:1vrep1}) to have quasi-parasupersymmetry of order $(3,2)$
as follows:
\begin{subequations}
\label{eq:qp1c8}
\begin{align}
2H_{0}=Q_{1}^{-}Q_{1}^{+}-2\bar{R}_{1},\quad 2H_{1}=Q_{1}^{+}Q_{1}^{-}
 -2\bar{R}_{1}=Q_{2}^{-}Q_{2}^{+}-2\bar{R}_{2},\\
2H_{2}=Q_{2}^{+}Q_{2}^{-}-2\bar{R}_{2}=Q_{3}^{-}Q_{3}^{+}-2\bar{R}_{3},
 \quad 2H_{3}=Q_{3}^{+}Q_{3}^{-}-2\bar{R}_{3},
\end{align}
\end{subequations}
where the new set of constants $\bar{R}_{k}$ ($k=1,2,3$) are introduced by
\begin{align}
3\bar{R}_{1}=2R_{1}-R_{2}-R_{3},\quad 3\bar{R}_{2}=-R_{1}+2R_{2}-R_{3},
 \quad 3\bar{R}_{3}=-R_{1}-R_{2}+2R_{3},
\label{eq:defbR}
\end{align}
and thus satisfy
\begin{align}
\bar{R}_{1}+\bar{R}_{2}+\bar{R}_{3}=0.
\end{align}
Hence, it is completely equivalent to the (ordinary) parasupersymmetric
system (\ref{eq:1vrep2}). It is evident that the same generalized $3$-fold
algebra (\ref{eq:g3falg}) also holds (with $R_{k}\to\bar{R}_{k}$).

We next consider quasi-parasupersymmetry of order $(3,3)$. We recall
the fact that we have already solved the conditions (\ref{eq:3psc2})--%
(\ref{eq:3psc3}) and (\ref{eq:3psc2'})--(\ref{eq:3psc3'}) to obtain
Eqs.~(\ref{eq:1vrep3}) and (\ref{eq:qp1c6}). Thus, the remaining
conditions to be satisfied are Eq.~(\ref{eq:3qpsc2}).
Substituting the expressions (\ref{eq:1vrep1}) and (\ref{eq:qp1c6}) into
the conditions (\ref{eq:3qpsc2}), we find that they are satisfied if
and only if
\begin{align}
2V_{0}=W_{3}^{2}-2W'_{1}-2W'_{2}-W'_{3}+\frac{2}{3}(R_{1}+R_{2}-2R_{3}),
\label{eq:qp1c9}\\
W''_{1}+3W''_{2}+2W''_{3}+2W_{1}W'_{1}+2W_{2}W'_{2}-4W_{3}W'_{3}=0,
\label{eq:qp1c10}
\end{align}
and
\begin{multline}
W'''_{2}+W'''_{3}-W_{1}W''_{2}-W_{1}W''_{3}+3W_{2}W''_{2}+2(W'_{2})^{2}
 +W_{2}W''_{3}-2W_{3}W''_{3}\\
-2(W'_{3})^{2}-2W_{1}W_{2}W'_{2}+2W_{1}W_{3}W'_{3}
 +2W_{2}^{2}W'_{2}-2W_{2}W_{3}W'_{3}=0.
\label{eq:qp1c11}
\end{multline}
It is easy to check that the condition (\ref{eq:1vrep1}) implies both
(\ref{eq:qp1c10}) and (\ref{eq:qp1c11}). Hence, the necessary and
sufficient condition for the system (\ref{eq:1vrep1}) to have
quasi-parasupersymmetry of order $(3,3)$ is as follows:
\begin{subequations}
\begin{align}
2H_{0}=Q_{1}^{-}Q_{1}^{+}-2\bar{R}_{1},\quad Q_{1}^{+}Q_{1}^{-}
 -2\bar{R}_{1}=Q_{2}^{-}Q_{2}^{+}-2\bar{R}_{2},\\
2H_{2}=Q_{2}^{+}Q_{2}^{-}-2\bar{R}_{2}=Q_{3}^{-}Q_{3}^{+}-2\bar{R}_{3},
 \quad 2H_{3}=Q_{3}^{+}Q_{3}^{-}-2\bar{R}_{3},
\end{align}
\end{subequations}
where $\bar{R}_{k}$ ($k=1,2,3$) are the same as Eq.~(\ref{eq:defbR}).
We note in particular that there are no constraints on the form of the
Hamiltonian $H_{1}$ in contrast to the cases of order $(3,1)$ and
$(3,2)$.

\subsection{Three-Variable Representation}
\label{ssec:exam3}

As another example, let us next consider a three-body system given by
\begin{align}
H_{k}&=-\frac{1}{2}\sum_{i=1}^{3}\frac{\del^{2}}{\del q_{i}^{2}}
 +V_{k}(q_{1},q_{2},q_{3}),\qquad
Q_{k}^{\pm}=\pm\frac{\del}{\del q_{k}}+W_{k}(q_{1},q_{2},q_{3}).
\label{eq:3vrep1}
\end{align}
The first-order intertwining relations (\ref{eq:3psc1}) and
(\ref{eq:3psc1'}) are satisfied if and only if
\begin{align}
W_{k}(q_{1},q_{2},q_{3})=W_{k}(q_{k})\quad(k=1,2,3),
\label{eq:3vrep2}
\end{align}
and
\begin{subequations}
\label{eq:3vrep3}
\begin{align}
2V_{0}(q_{1},q_{2},q_{3})&=W_{1}(q_{1})^{2}-W'_{1}(q_{1})
 +f_{1}(q_{2},q_{3}),\\
2V_{1}(q_{1},q_{2},q_{3})&=W_{1}(q_{1})^{2}+W'_{1}(q_{1})
 +f_{1}(q_{2},q_{3})\notag\\
&=W_{2}(q_{2})^{2}-W'_{2}(q_{2})+f_{2}(q_{1},q_{3}),
\label{eq:3vrep32}\\
2V_{2}(q_{1},q_{2},q_{3})&=W_{2}(q_{2})^{2}+W'_{2}(q_{2})
 +f_{2}(q_{1},q_{3})\notag\\
&=W_{3}(q_{3})^{2}-W'_{3}(q_{3})+f_{3}(q_{1},q_{2}),
\label{eq:3vrep33}\\
2V_{3}(q_{1},q_{2},q_{3})&=W_{3}(q_{3})^{2}+W'_{3}(q_{3})
 +f_{3}(q_{1},q_{2}),
\end{align}
\end{subequations}
where $f_{k}$ ($k=1,2,3$) are certain functions of two variables.
The two equalities in Eqs.~(\ref{eq:3vrep32}) and (\ref{eq:3vrep33})
are compatible if and only if
\begin{subequations}
\label{eq:3vrep4}
\begin{align}
f_{1}(q_{2},q_{3})&=W_{2}(q_{2})^{2}-W'_{2}(q_{2})
 +W_{3}(q_{3})^{2}-W'_{3}(q_{3})-2R,\\
f_{2}(q_{1},q_{3})&=W_{1}(q_{1})^{2}+W'_{1}(q_{1})
 +W_{3}(q_{3})^{2}-W'_{3}(q_{3})-2R,\\
f_{3}(q_{1},q_{2})&=W_{1}(q_{1})^{2}+W'_{1}(q_{1})
 +W_{2}(q_{2})^{2}+W'_{2}(q_{2})-2R,
\end{align}
\end{subequations}
where $R$ is a constant. Substituting (\ref{eq:3vrep4}) into
(\ref{eq:3vrep3}), we have
\begin{subequations}
\label{eq:3vrep5}
\begin{align}
2H_{0}&=Q_{1}^{-}Q_{1}^{+}+Q_{2}^{-}Q_{2}^{+}+Q_{3}^{-}Q_{3}^{+}-2R,&
2H_{1}&=Q_{1}^{+}Q_{1}^{-}+Q_{2}^{-}Q_{2}^{+}+Q_{3}^{-}Q_{3}^{+}-2R,\\
2H_{2}&=Q_{1}^{+}Q_{1}^{-}+Q_{2}^{+}Q_{2}^{-}+Q_{3}^{-}Q_{3}^{+}-2R,&
2H_{3}&=Q_{1}^{+}Q_{1}^{-}+Q_{2}^{+}Q_{2}^{-}+Q_{3}^{+}Q_{3}^{-}-2R.
\end{align}
\end{subequations}
Finally, substituting (\ref{eq:3vrep5}) into the conditions
(\ref{eq:3psc2}) or (\ref{eq:3psc2'}), we find that they are
satisfied if and only if
\begin{align}
C_{3}=2,\qquad R=0.
\label{eq:3vrep6}
\end{align}
In this case, the Hamiltonian $H_{2}$ satisfies Eq.~(\ref{eq:3psc4})
and thus the remaining conditions (\ref{eq:3psc3}) and (\ref{eq:3psc3'})
are automatically fulfilled. Hence, the necessary and sufficient
conditions for the system (\ref{eq:3vrep1}) to have parasupersymmetry of
order $3$ are given by Eqs.~(\ref{eq:3vrep2}), (\ref{eq:3vrep5}), and
(\ref{eq:3vrep6}). We can easily see that there are three independent
supersymmetries folded in the system (\ref{eq:3vrep5}). We note that
the parasupersymmetric system $\bH$ in this case is given by
\begin{align}
2\bH=&\, (Q_{1}^{-}Q_{1}^{+}+Q_{2}^{-}Q_{2}^{+}+Q_{3}^{-}Q_{3}^{+})
 \Pi_{0}+(Q_{1}^{+}Q_{1}^{-}+Q_{2}^{-}Q_{2}^{+}+Q_{3}^{-}Q_{3}^{+})
 \Pi_{1}\notag\\
&\, +(Q_{1}^{+}Q_{1}^{-}+Q_{2}^{+}Q_{2}^{-}+Q_{3}^{-}Q_{3}^{+})\Pi_{2}
 +(Q_{1}^{+}Q_{1}^{-}+Q_{2}^{+}Q_{2}^{-}+Q_{3}^{+}Q_{3}^{-})\Pi_{3}.
\end{align}
Comparing it with Eqs.~(\ref{eq:mono1})--(\ref{eq:mono6}), we observe
that any function of $\bH$ cannot be expressed as a function of
$\bQ^{-}$ and $\bQ^{+}$, in contrast to the previous one-body case.
The situation is analogous to that of the two-body second-order
parasupersymmetric systems in Ref.~\cite{Ta07a}.

Next, we shall examine quasi-parasupersymmetry in the system given
by Eq.~(\ref{eq:3vrep1}). For the purpose, we first solve the conditions
(\ref{eq:3psc2}) and (\ref{eq:3psc2'}). Substituting the expression
(\ref{eq:3vrep1}) into the conditions (\ref{eq:3psc2}) or
(\ref{eq:3psc2'}), we can easily find that they are satisfied if and
only if $C_{3}=2$ and
\begin{align}
2V_{2}=W_{1}^{2}+(\del_{1}W_{1})+W_{2}^{2}+(\del_{2}W_{2})
 +W_{3}^{2}-(\del_{3}W_{3}),
\label{eq:qp3c1}\\
(\del_{1}W_{2})=0,\qquad (\del_{1}\del_{2}W_{1})
 +2W_{1}(\del_{2}W_{1})=0,
\label{eq:qp3c2}
\end{align}
where $(\del_{i}f)=\del f/\del q_{i}$.
Similarly, if we substitute Eq.~(\ref{eq:3vrep1}) into the conditions
(\ref{eq:3psc3}) or (\ref{eq:3psc3'}) with $C_{3}=2$, we find
\begin{align}
2V_{3}=W_{1}^{2}+(\del_{1}W_{1})+W_{2}^{2}+(\del_{2}W_{2})
 +W_{3}^{2}+(\del_{3}W_{3}),
\label{eq:qp3c3}\\
(\del_{1}W_{2})=(\del_{1}W_{3})=(\del_{2}W_{3})=0,\qquad
(\del_{1}\del_{2}W_{1})+2W_{1}(\del_{2}W_{1})=0,
\label{eq:qp3c4}\\
(\del_{1}\del_{3}W_{1})+2W_{1}(\del_{3}W_{1})+(\del_{2}\del_{3}W_{2})
 +2W_{2}(\del_{3}W_{2})=0.
\label{eq:qp3c5}
\end{align}
We now first consider quasi-parasupersymmetry of order $(3,2)$. The
remaining conditions to be satisfied are Eqs.~(\ref{eq:3qpsc1}).
Substituting the expressions (\ref{eq:3vrep1}) into the conditions
(\ref{eq:3qpsc1}) and using Eqs.~(\ref{eq:qp3c1})--(\ref{eq:qp3c5}),
we find that they are satisfied if and only if
\begin{align}
2V_{0}=W_{1}^{2}-(\del_{1}W_{1})+W_{2}^{2}-(\del_{2}W_{2})
 +W_{3}^{2}-(\del_{3}W_{3}),
\label{eq:qp3c6}\\
2V_{1}=W_{1}^{2}+(\del_{1}W_{1})+W_{2}^{2}-(\del_{2}W_{2})
 +W_{3}^{2}-(\del_{3}W_{3}),
\label{eq:qp3c7}\\
(\del_{2}W_{1})=(\del_{3}W_{1})=(\del_{3}W_{2})=0.
\label{eq:qp3c8}
\end{align}
It is evident that the conditions (\ref{eq:qp3c2}), (\ref{eq:qp3c4}),
(\ref{eq:qp3c5}), and (\ref{eq:qp3c8}) are altogether equivalent to
Eq.~(\ref{eq:3vrep2}). In this case, the formulas (\ref{eq:qp3c1}),
(\ref{eq:qp3c3}), (\ref{eq:qp3c6}), and (\ref{eq:qp3c7}) are identical
to Eqs.~(\ref{eq:3vrep5}) with $R=0$. Hence, for the system given by
Eq.~(\ref{eq:3vrep1}) quasi-parasupersymmetry of order $(3,2)$ is 
again completely equivalent to the (ordinary) parasupersymmetry.

We next consider quasi-parasupersymmetry of order $(3,3)$. We have
already solved the conditions (\ref{eq:3psc2})--(\ref{eq:3psc3})
and (\ref{eq:3psc2'})--(\ref{eq:3psc3'}) to obtain
Eqs.~(\ref{eq:qp3c1})--(\ref{eq:qp3c5}). Thus, the remaining
conditions to be satisfied are Eq.~(\ref{eq:3qpsc2}).
Substituting the expressions (\ref{eq:3vrep1}) into the conditions
(\ref{eq:3qpsc2}) and applying Eqs.~(\ref{eq:qp3c1})--(\ref{eq:qp3c5}),
we find that they are satisfied if and only if
\begin{align}
2V_{0}=W_{1}^{2}-(\del_{1}W_{1})+W_{2}^{2}-(\del_{2}W_{2})
 +W_{3}^{2}-(\del_{3}W_{3}),
\label{eq:qp3c9}\\
(\del_{2}W_{1})=(\del_{3}W_{1})=(\del_{3}W_{2})=0.
\label{eq:qp3c10}
\end{align}
Hence, for the three-body system (\ref{eq:3vrep1}) parasupersymmetry
of order $(3,3)$ is identical to those of order $(3,1)$ and $(3,2)$
except for the fact that there are no constraints on the form of
the Hamiltonian $H_{1}$, as in the case of the one-body system in
Section~\ref{ssec:exam1}.

\section{Discussion and Summary}
\label{sec:discus}

In this article, we have investigated third-order parafermionic
algebra and parasupersymmetric quantum systems based on the general
formalism in our previous work. We have found that the postulates
of the formalism work well also in the case of third order and enable
us to construct systematically the parafermionic algebra and the
multiplication law. We have constructed the two different third-order
parasupersymmetric quantum systems, the one consists of one-body
Hamiltonians and the other consists of three-body ones. They are
respectively natural generalizations of the one-body and two-body
second-order systems in our previous article \cite{Ta07a}. We have
also investigated quasi-parasupersymmetry in those two systems and
found that the order $(3,2)$ cases are equivalent to the order $(3,1)$,
namely, the ordinary third-order parasupersymmetric cases, while
the order $(3,3)$ cases are realized with the less restrictive conditions
by dropping the constraint on the component Hamiltonian $H_{1}$.

Although most of the features in the third-order case we have found
in this article have strong resemblance to those in the second-order
case, there are some novel features in the former which do not appear
in the latter. One is the splitting of the anti-commutation relation
(\ref{eq:3alg1}) into the odd- and even-degree parts, namely,
Eq.~(\ref{eq:alg3d}). As a result, the part of the fundamental elements,
$\{I,(\psi^{-})^{2},(\psi^{+})^{2}\}$ behaves as an ordinary fermionic
system. From this result, we conjecture that a similar decomposition
would take place in arbitrary odd-order parafermionic algebra and in
particular the part $\{I,(\psi^{-})^{p},(\psi^{+})^{p}\}$ in
$(2p-1)$th-order for all $p=1,2,3,\ldots$ would also behave as
a fermionic system. It is also interesting to see whether some
decomposition of the relation (\ref{eq:postu}) takes place in even
$(2p)$th-order cases for $p\geq 2$.

Another new feature in the third-order is the emergence of the
generalized parafermionic statistics characterized by the quadrilinear
relations (\ref{eq:quadri}). From this result, we conjecture that
in our formalism parafermionic operators of order $p$ are characterized
by $(p+1)$-tuple linear relations. We have not appreciated whether
such a generalized statistics is compatible with other physical
requirements, especially in view of the canonical formulation of quantum
theory (cf. Refs.~\cite{Gr52,OK82,GM65}). But we hope it could provide
a new possibility in both physical and mathematical studies. We also
note that the expected $(p+1)$-tuple linear relations in $p$th-order
could be a clue to improving the BDCF formalism, that is, how to
modify the left-hand sides of Eqs.~(\ref{eq:pfsc4}).

We have found that the quasi-parasupersymmetric quantum systems of
order $(3,2)$ obtained in this article are identical to the
corresponding parasupersymmetric ones. This result, together with
the one that all the quasi-parasupersymmetric quantum systems of order
$(2,2)$ obtained in Ref.~\cite{Ta07a} are also identical with the
corresponding parasupersymmetric ones, indicates that the conditions
of parasupersymmetry is in fact too strong, at least, for the second-
and third-order cases. On the other hand, the peculiar feature of the
order $(3,3)$ cases where only the component Hamiltonian $H_{1}$ has
no restriction stems from the fact that the non-linear constraints
(\ref{eq:parac}) impose the restrictions only on $H_{p-1}$ and
$H_{p}$ among the component Hamiltonians $H_{k}$ ($k=0,\dots,p$).
In other words, the conditions (\ref{eq:parac}) unnaturally violate
equality of component Hamiltonians. These observations may suggest
that there is a more suitable definition or formulation of
parasupersymmetry.

The discovery of generalized $\cN$-fold superalgebra in the one-body
parasupersymmetric quantum systems of generalized Rubakov--Spiridonov
type for second- and third-order clearly indicates that it also exists
in this type of models for arbitrary order $p$. Indeed, the component
Hamiltonians $H_{0}$ and $H_{p}$ in this type are always a one-body
$\cN$-fold supersymmetric pair ($\cN=p$) with respect to the
components of the `$\cN$-fold' supercharges $(\bQ^{\pm})^{p}$ and thus
in the subsector $\fF\times(\sV_{p}^{(0)}\dotplus\sV_{p}^{(p)})$
the triple $(\bH,(\bQ^{-})^{p},(\bQ^{+})^{p})$ must satisfy an $\cN$-fold
superalgebra. Hence, in the whole space $\fF\times\sV_{p}$ we can
naturally expect some generalized form of it. We also note that
the absence of any additional algebraic relation in the two- and
three-body parasupersymmetric quantum systems would explain the reason
why there have been no satisfactory formulation of $\cN$-fold
supersymmetry in quantum many-body systems solely in terms of
commutators and anti-commutators. The parasuperalgebra
(\ref{eq:pfsc1})--(\ref{eq:pfsc3}) thus can be an alternative
framework to realize higher-order intertwining relations among
many-body Hamiltonians.

Gathering altogether the knowledge so far obtained in the second- and
third-order cases, we hope we would be able to report some inductive
study on higher-order cases in the near future.

\begin{acknowledgments}
 This work was partially supported by the National Cheng-Kung
 University under the grant No. OUA:95-3-2-071.
\end{acknowledgments}


\bibliography{refsels}

\begin{thebibliography}{10}
\expandafter\ifx\csname url\endcsname\relax
  \def\url#1{{\tt #1}}\fi
\expandafter\ifx\csname urlprefix\endcsname\relax\def\urlprefix{URL }\fi
\providecommand{\eprint}[2][]{\url{#2}}

\bibitem{Ta07a}
T.~Tanaka, Ann. Phys. {(2007), doi:10.1016/j.aop.2006.11.009, in press},
  \eprint{hep-th/0610311}.

\bibitem{FV91}
R.~Floreanini and L.~Vinet, Phys. Rev. D 44 (1991) 3851.

\bibitem{BD93b}
J.~Beckers and N.~Debergh, Int. J. Mod. Phys. A 8 (1993) 5041.

\bibitem{Sm03}
A.~V. Smilga, Phys. Lett. B 585 (2004) 173.
\newblock \eprint{hep-th/0311023}.

\bibitem{AST01b}
H.~Aoyama, M.~Sato, and T.~Tanaka, Nucl. Phys. B 619 (2001) 105.
\newblock \eprint{quant-ph/0106037}.

\bibitem{RS88}
V.~A. Rubakov and V.~P. Spiridonov, Mod. Phys. Lett. A 3 (1988) 1337.

\bibitem{Kh92}
A.~Khare, J. Phys. A: Math. Gen. 25 (1992) L749.

\bibitem{Kh93}
A.~Khare, J. Math. Phys. 34 (1993) 1277.

\bibitem{To92}
M.~Tomiya, J. Phys. A: Math. Gen. 25 (1992) 4699.

\bibitem{BD90a}
J.~Beckers and N.~Debergh, Nucl. Phys. B 340 (1990) 767.

\bibitem{CF03}
A.~Chenaghlou and H.~Fakhri, Int. J. Mod. Phys. A 18 (2003) 939.
\newblock \eprint{hep-th/0203240}.

\bibitem{Ta06b}
T.~Tanaka, J. Phys. A: Math. Gen. 39 (2006) L369.
\newblock \eprint{hep-th/0603096}.

\bibitem{Ta06d}
T.~Tanaka, J. Phys. A: Math. Gen. 39 (2006) 14175.
\newblock \eprint{hep-th/0605035}.

\bibitem{Wi81}
E.~Witten, Nucl. Phys. B 188 (1981) 513.

\bibitem{Ta03a}
T.~Tanaka, Nucl. Phys. B 662 (2003) 413.
\newblock \eprint{hep-th/0212276}.

\bibitem{Mo96a}
A.~Mostafazadeh, Int. J. Mod. Phys. A 11 (1996) 1057.
\newblock \eprint{hep-th/9410180}.

\bibitem{Mo97}
A.~Mostafazadeh, Int. J. Mod. Phys. A 12 (1997) 2725.
\newblock \eprint{hep-th/9603163}.

\bibitem{Gr52}
H.~S. Green, Phys. Rev. 90 (1952) 270.

\bibitem{OK82}
Y.~Ohnuki and S.~Kamefuchi, {Q}uantum {F}ield {T}heory and {P}arastatistics
  (Springer, New York, 1982).

\bibitem{DV90a}
S.~Durand and L.~Vinet, Phys. Lett. A 146 (1990) 299.

\bibitem{GM65}
O.~W. Greenberg and A.~M.~L. Messiah, Phys. Rev. 138 (1965) B1155.

\end{thebibliography}
\bibliographystyle{npb}

\end{document}